\documentclass[11pt, a4paper]{article}
\usepackage[utf8]{inputenc}
\usepackage[left=2cm, right=2cm, bottom=2cm, top=2cm]{geometry}
\usepackage{xcolor}
\usepackage{bm}
\usepackage{amsmath, amsfonts}
\usepackage{siunitx}
\DeclareSIUnit{\angstrom}{\textup{\AA}}
\usepackage[hyphens]{url}
\usepackage{booktabs}
\usepackage{graphicx}
\usepackage{caption}
\usepackage{csquotes}
\usepackage{empheq}
\usepackage{xparse}
\definecolor{amethyst}{rgb}{0.6, 0.4, 0.8}
\definecolor{revcol}{rgb}{0,0,1}
\usepackage[square, numbers, comma, sort&compress]{natbib}
\usepackage[colorlinks, citecolor=blue, linkcolor=teal, urlcolor=amethyst]{hyperref}
\usepackage{float}

\usepackage[capitalise,noabbrev,nameinlink]{cleveref}
\crefname{sfig}{Supplementary Figure}{Supplementary Figures}
\crefname{stab}{Supplementary Table}{Supplementary Tables}
\usepackage{authblk}

\usepackage[square, numbers, comma, sort&compress]{natbib}

\title{Promotion of cooperation in deme-structured populations with growth-merging dynamics}
\author{Damien Ribière\textsuperscript{1,2}, Alia Abbara\textsuperscript{1,2,$\dagger$}, Anne-Florence Bitbol\textsuperscript{1,2,*}}
\affil{\textbf{1} Institute of Bioengineering, School of Life Sciences, École Polytechnique Fédérale de Lausanne (EPFL), CH-1015 Lausanne, Switzerland\\
\textbf{2} SIB Swiss Institute of Bioinformatics, CH-1015 Lausanne, Switzerland\\
$\dagger$ Current address: Sorbonne Université, CNRS, Institut de
Biologie Paris-Seine (IBPS), Laboratoire Jean Perrin (LJP), Paris, France\\
* Corresponding author: \href{mailto:anne-florence.bitbol@epfl.ch}{anne-florence.bitbol@epfl.ch}}
\date{}

\begin{document}

\maketitle

\begin{abstract}
The spatial structure of populations may promote the emergence and maintenance of cooperation. Cooperation in the prisoner's dilemma is favored under specific update rules in evolutionary graph theory models with one individual per node of a graph, but this effect vanishes in models with well-mixed demes connected by migrations under soft selection (each deme contributing independently of its productivity). In contrast, experiments and models involving cycles of growth, merging and dilution have shown that spatial structure can favor cooperation. Here, we reconcile these findings by studying deme-structured populations under growth-merging-dilution dynamics, corresponding to a clique (fully connected graph) under hard selection (each deme contributing in proportion to its productivity). We obtain analytical conditions for the cooperator fraction to increase during deterministic logistic growth, and to increase on average under dilution-growth-merging cycles, in the weak selection regime, where fitness differences between genotypes are small. Furthermore, we analytically express the fixation probability of cooperators under weak selection, yielding a criterion for cooperators to have a higher fixation probability than neutral mutants. Finally, numerical simulations show that stochastic growth further promotes cooperation. Overall, hard selection is essential for cooperation to be promoted in deme-structured populations.
\end{abstract}

\section*{Introduction}

Understanding the emergence and maintenance of cooperation is an important question in biology. As cooperative behavior benefits other individuals but entails a cost for the cooperator individual, non-cooperators tend to be advantaged by natural selection. Nevertheless, cooperation is often observed in nature, from microorganisms to animals~\cite{axelrod81,Wingreen06,celiker2013cellular,dugatkin1997cooperation,chase1980cooperative}. 
This apparent paradox is known as the dilemma of cooperation \cite{Review_2019}. 
In particular, bacteria interact by secreting various molecules, some of which are public goods, i.e.\ shared resources that bring a metabolic cost for the producers, but benefit all other bacteria in the population, including non-producers (cheaters)~\cite{Borenstein13,Review_2019}. Different mechanisms may explain the evolution of cooperation under some conditions. One of them is kin selection~\cite{Hamilton64a,Hamilton64b,griffin2004cooperation,belcher2022kin}: if individuals that are genetically related to cooperators most benefit from cooperation, then it can be selected despite its cost. Another one is spatial structure, which has been investigated theoretically~\cite{ohtsuki2006evolutionary,Nowak_2006,Cremer_2010,Cremer_2011,Cremer_2012,Melbinger_2015,Review_2019,czuppon_spatial_2017,Arenzon23} and experimentally~\cite{Simpson_2009,Simpson_2010,Becker_2018}. Since genetically related individuals are often closely located in spatially structured populations, the ability of spatial structure to enhance cooperation is related to kin selection~\cite{Kay20}. 

In evolutionary graph theory, a simple and minimal model of cooperation in spatially structured populations is the prisoner's dilemma game~\cite{Turner99,doebeli2005,gokhale_evolutionary_2014}, played on graphs with one individual per node~\cite{lieberman2005evolutionary}. It was shown that cooperation can be favored by natural selection in the prisoner's dilemma game in some graph structures \cite{Nowak_2006,ohtsuki2006evolutionary}: a cooperator is then more likely to take over the structured population than a neutral mutant. However, these results strongly depend on the update rule that specifies how individuals divide, die, and replace each other (e.g.\ death-birth or birth-death). Recently, we developed general models of structured populations \cite{Marrec_2021, Abbara_2023}, where each node of the graph comprises a well-mixed deme (i.e.\ subpopulation) instead of a single individual. Our models do not involve choosing an individual-level update rule connecting the birth, death and migration events of individuals, which impacts fixation probabilities~\cite{Houchmandzadeh11,yagoobi2021fixation,yagoobi2023}. Such a level of description is difficult to justify for microbes, unlike deme-level protocols that can be set experimentally~\cite{Chakraborty23}. The deme-level demographic scheme nonetheless shapes evolutionary outcomes, as we will show here. In Ref.~\cite{Moawad_2023}, we studied the evolution of cooperation in these models under soft selection, namely when each deme contributes a number of individuals independent of its composition. We showed that cooperation is not favored by spatial structure in this case: a cooperator is less likely to take over the structured population than a neutral mutant~\cite{Moawad_2023}. However, previous studies that considered populations undergoing cycles of growth in subpopulations, merging of all subpopulations and bottlenecks yielding new subpopulations demonstrated that the total fraction of cooperators can increase in time \cite{Simpson_2009,Simpson_2010,Becker_2018,Cremer_2010,Cremer_2011,Cremer_2012,Melbinger_2015,Review_2019}. This occurs due to Simpson's paradox \cite{Simpson_2009}: if there is sufficient heterogeneity of composition between subpopulations, the fraction of cooperators can increase overall, despite its decrease in each separate subpopulation which results from the individual cost of cooperation, because subpopulations with more cooperators grow faster than others. 

Here, we reconcile the results of Ref.~\cite{Moawad_2023} with those of Refs.~\cite{Simpson_2009,Simpson_2010,Becker_2018,Cremer_2010,Cremer_2011,Cremer_2012,Melbinger_2015,Review_2019}. As in Ref.~\cite{Moawad_2023}, we use our general model of structured populations \cite{Abbara_2023}, which accounts for cycles of growth and dilution-migration, and we consider individuals that interact via a prisoner's dilemma game played in each deme of a structured population. However, we consider that all demes are merged before a new bottleneck state is formed for each deme by sampling from the merged population, as in Refs.~\cite{Cremer_2012,Melbinger_2015}.  Since demes contribute to the next bottleneck state proportionally to their grown sizes, this model implements a simple case of hard selection, where each deme contributes a number of individuals proportional to its productivity~\cite{wallace1975hard,christiansen1975hard,ravigne2009live}. Given the symmetry between all demes, this structured population can be viewed as a clique graph (or fully connected graph) with hard selection. This stands in contrast with Ref.~\cite{Moawad_2023}, where soft selection was assumed.

We analytically determine the time evolution of the cooperator fraction during a deterministic logistic growth phase, in the weak selection regime, where the fitness differences between genotypes are small (a distinct notion from the hard versus soft distinction above, which regards how demes contribute). We interpret our result as a multi-level Price equation \cite{Price_1972,Jeler_2018}, featuring a tradeoff between within-deme dynamics, where cooperators are disadvantaged, and between-deme dynamics, where demes with more cooperators are advantaged. We determine a criterion for cooperator fraction to increase during the growth phase. 
Next, we analytically express the expectation value of the variation of the overall cooperator fraction in the structured population, over a full cycle of dilution-growth-merging under weak selection. This allows us to define an effective fitness parameter that determines whether the overall cooperator fraction is expected to grow on average over the dilution-growth-merging dynamics. An important condition for this to happen is that deme bottleneck sizes should be small enough. Then, we analytically determine the fixation probability of cooperators, assuming weak selection but without assuming large demes. We obtain a condition for the fixation of a cooperator to be more favorable than that of a neutral mutant. Finally, using numerical simulations, we explore the impact of stochastic growth on our results, and find that it facilitates the promotion of cooperation.

\section*{Model and Methods}

\subsection*{Spatially structured population with cycles of growth, merging and dilution}

We model spatially structured populations on a fully connected graph (or clique) with $D$ nodes, each one comprising a well-mixed subpopulation, also known as a deme. We consider two possible types of individuals: cooperators and defectors (or cheaters). As we are interested in the emergence of cooperation, we denote cooperators by $M$ (mutants) and defectors by $W$ (wild-type), and their respective fitnesses by $f_M$ and $f_W$. Fitnesses are understood as their maximal growth rates in the exponential phase. 

In a recent study \cite{Moawad_2023}, we investigated the evolution of cooperation in deme-structured populations on graphs, using models we introduced in \cite{Marrec_2021,Abbara_2023}. We assumed that demes are connected by migrations under soft selection (i.e.\ migrations are not affected by deme size). This gave rise to a separation of scales where cooperation had an impact only within demes, and not at the scale of the population. As a result, cooperation was not advantaged by spatial structure. 
Here, we extend the model of deme-structured populations on graphs from Ref. \cite{Abbara_2023} to a specific case of hard selection, where migrations are proportional to the size of the deme of origin after growth, and we investigate the evolution of cooperation in this case. Specifically, the structured population undergoes three-step cycles of deterministic logistic growth with carrying capacity $K$ within each deme, merging of all demes and dilution and formation of new demes, as follows. 

\subsection*{Growth in each deme}

Starting from a bottleneck with $B$ individuals, each deme undergoes a logistic growth phase during time $t$, with a carrying capacity $K \gg 1$. To describe cooperation in a minimal way, we model interactions between individuals in a deme with the prisoner's dilemma game, as in \cite{Moawad_2023}, and inspired by evolutionary game theory models on graphs with one individual per node \cite{Nowak_2006,OHTSUKI200686}. The payoff matrix reads
\begin{equation}
P=\left(\begin{array}{cc}b-c & -c \\ b & 0\end{array}\right) \; ,
\end{equation}
where $b>c>0$. Here, $b$ models the benefit brought by interacting with a cooperator, while $c$ models the individual cost of cooperation for a cooperator. If a cooperator interacts with a defector, the cooperator receives a payoff $-c<0$ whereas the cheater receives a payoff $b>c>0$. If two cooperators interact, they both receive a payoff $b-c$. Finally, if two defectors interact, the payoff is zero.

In a well-mixed deme, each individual interacts with all other individuals. Denoting by $k$ the number of cooperators, $l$ the numbers of defectors in a deme and $N= k + l$ the deme size, the total payoff of any individual of type $M$ or $W$ is
\begin{equation}
    \begin{cases}
    \pi_M(k, N) & = (b-c) \dfrac{k-1}{N-1}-c \dfrac{N-k}{N-1}\,, \\
    \pi_W(k, N) & = b \dfrac{k}{N-1} \; .
    \end{cases}
\end{equation}
We assume that an individual's fitness depends linearly on the payoff it receives, such that the frequency-dependent fitnesses $f_M(k, N)$ for a cooperator and $f_W (k, N)$ for a defector are given by
\begin{equation}
    \begin{cases}
    f_M(k, N) & = 1-w+w \pi_M(k, N)\,, \\
    f_W(k, N) & = 1-w+w \pi_W(k, N) \,,
    \end{cases}
    \label{formules}
\end{equation}
with $w$ the intensity of selection. We will interpret fitness as growth rate in the exponential phase, and focus on the weak selection regime $w \ll 1$.

For $\tau \in [0,t]$, the numbers of cooperators and defectors satisfy the following ordinary differential equations
\begin{equation}
    \begin{cases}
    \dfrac{dk}{d\tau} = kf_M\left(1 - \dfrac{k+l}{K}\right)\,, \\
    \dfrac{dl}{d\tau} = lf_W\left(1 -\dfrac{k+l}{K}\right) \; . 
    \end{cases}
    \label{evolutions}
\end{equation}
Rewriting Eq.~\eqref{evolutions} in terms of deme size $N$ and of cooperator fraction $x=k/N$, leads to 
\begin{equation}
    \begin{cases}
    \dfrac{dN}{d\tau} = \left[1-w+w(b-c)x\right]N \left(1 -\dfrac{N}{K}\right) \; ,\\
       \dfrac{dx}{d\tau} = -wx(1-x)\left(\dfrac{b}{N-1}+c\right)\left(1 - \dfrac{N}{K}\right)\,.
\end{cases}
    \label{equation_diff}
\end{equation}
Note that $dx/d\tau\leq 0$ since $b>c>0$. Hence, in each deme, the fraction of cooperators decreases during the growth phase. Also note that the carrying capacity $K$ does not depend on the cooperator fraction: if growth time is long enough, all demes reach the same size $K$.

In the weak selection regime $w \ll 1$, the solution of this system of non-linear ordinary differential equations can be expanded as (See Supplement Section \ref{1.1} for details)
\begin{equation}
    \begin{cases}
    N(\tau) &= 
    \dfrac{K}{1+\frac{K-B}{B}e^{-\tau}} \left(1 + w  \dfrac{\tau[(b-c)x_0-1]}{1+\frac{B}{K-B}e^\tau}\right) + \mathcal{O}(w^2) \,, \\
    x(\tau) &= x_0-w x_0(1-x_0)\left[\tau(c-b)+b\log\left(\dfrac{B(K-1)e^{\tau}+B-K}{K(B-1)}\right)-c\log\left(\dfrac{B(e^{\tau}-1)+K}{K}\right)\right] \\
    & + \mathcal{O}(w^2) \,.
    \end{cases}
    \label{approximation}
\end{equation}
where $x_0$ denotes the initial proportion of cooperators. While these equations hold within a deme, we will also consider the overall fraction of cooperators in the structured population, denoted by $X(\tau)$, which is the average of cooperator fractions in each deme, weighted by their sizes.

\subsection*{Merging and dilution modeling hard selection}

\begin{figure}[H]
    \centering
    \includegraphics[width=0.6\textwidth]{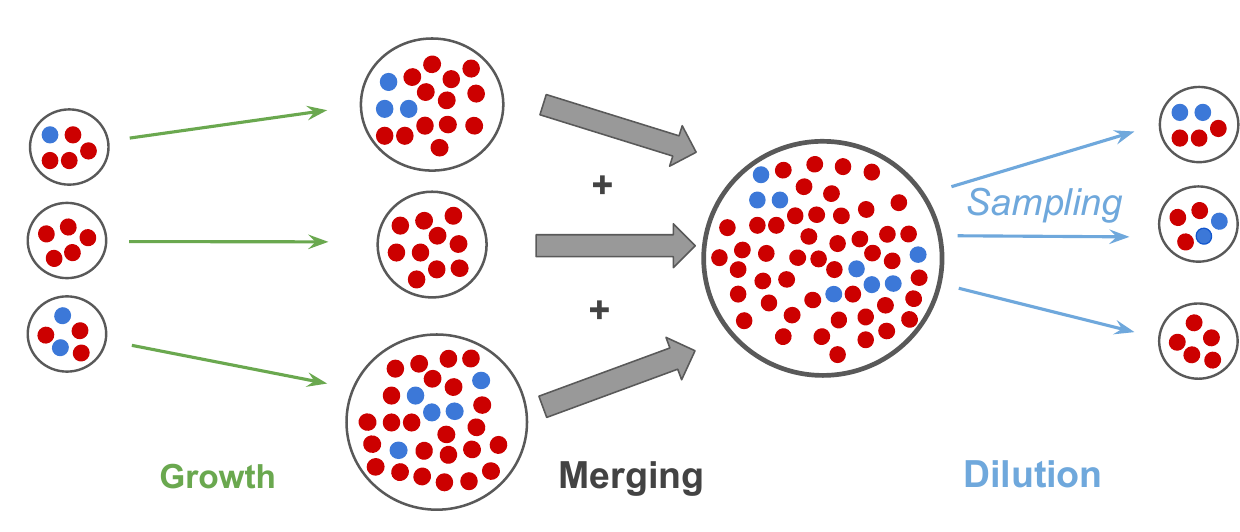}
    \caption{\textbf{Representation of one full cycle of growth-merging-dilution of the model for $D=3$ demes.} Blue markers represent cooperators, while red markers represent defectors. The initial state of the demes is shown on the left. The demes first undergo a local growth phase, followed by a merging where all individuals are put together. Then, the dilution step consists in three binomial samplings that determine the new bottleneck compositions of the three demes.}
    \label{Model}
\end{figure}

After the growth phase, we assume that all demes are merged to form one well-mixed population, and then new demes are formed by sampling from this well-mixed population, resulting in a new bottleneck state, see Fig. \ref{Model}. This process was previously considered in Refs.~\cite{Simpson_2009,Cremer_2010}. 
Specifically, after merging, the system is brought back to a new bottleneck state of $D$ demes, each of size $B$ by a dilution step, where each deme is formed by sampling $B$ individuals from the merged population. As the grown population is assumed to be large ($K \gg B$), we employ sampling with replacement. Hence, the new bottleneck state of each deme is formed by a binomial law with $B$ trials and a probability $X^\prime=X(t)$ of sampling a mutant. 

This merging and dilution process can be viewed as a particular case of migrations with hard selection in a clique. Selection is hard because each deme contributes to the next bottleneck state proportionally to its size after growth. To make the link explicit, consider a migration step on a clique graph of grown demes, followed by a local binomial sampling in each of them. In a clique, migrations occur with equal probability from one deme to another one. Here, we assume that migrations from each deme to itself also have the same probability, yielding a probability $m_{ij}=1/D$ for an individual in deme $i$ to migrate to deme $j$, for all $i$ and $j$. 
Let us denote by $N_i^\prime = N_i(t)$ (resp.\ $x_i^\prime = x_i(t)$) the size of deme $i$ (resp.\ the fraction of cooperators in that deme) after the growth phase. Assuming that grown demes are large enough for all fluctuations to be neglected, during the migration step, $N'_i m_{ij}=N'_i/D$ individuals, including $N'_i x'_i m_{ij}=N'_i x'_i/D$ mutants, go from deme $i$ to deme $j$. The fact that these numbers are proportional to the grown size $N_i^\prime$ of the deme of origin $i$ illustrates that the process comprises hard selection. After this migration step, all demes have the same size $\sum_i N'_i/D$ and the same mutant fraction $X^\prime=\sum_iN'_ix'_i/\sum_iN'_i$. 
Then, binomial sampling is performed locally in each deme to retain only $B$ individuals per deme, with a probability $X^\prime$ of sampling a mutant. Note that this corresponds to the usual sampling method in structured Wright-Fisher models~\cite{BURDEN201870}, and is one of the sampling schemes  considered in Ref.~\cite{Abbara_2023} (under the name ``local binomial'').

\section*{Results}

\subsection*{Cooperation can be transiently promoted on the graph during a growth phase}

We first ask whether the overall cooperator fraction can increase during a growth phase. To address this question, starting from Eq.~\eqref{approximation}, we express the excess overall cooperator fraction $\Delta X(\tau)= X(\tau)- X(0)$ in the population during growth (for $\tau \in[0, t]$), see Supplement Section \ref{1.2.1}. We obtain, in the weak fitness intensity regime $w \ll 1$,
\begin{align}
    \Delta X(\tau) &= w\left[\Delta X_a (\tau) - \Delta X_b (\tau)\right] + \mathcal{O}\left(w^2\right) \label{Delta_X_Final} ,\;\;\textrm{with}\\
    \Delta X_a (\tau) &= \mathrm{Var}(x_i) \frac{\tau(b-c)}{1+\frac{B}{K-B}e^\tau} \; ,\;\;\textrm{and}\label{Delta_Xa}\\
    \Delta X_b(\tau) &= \left \langle x_i(1-x_i)\right \rangle \left[\tau(c-b)+b\log\left(\frac{B(K-1)e^{\tau}+B-K}{K(B-1)}\right)-c \log\left(\frac{B(e^{\tau}-1)+K}{K}\right)\right] \; ,\label{Delta_Xb}
\end{align}
where $x_i$ denotes the initial fraction of cooperators in deme $i$, for $1 \le i \le D$. Here, averages over demes are denoted by $\langle . \rangle = (1/D)\sum_i (.)$, and the variance in the initial cooperator fraction across demes is $\mathrm{Var}(x_{i}) = \langle x_{i}^2 \rangle - \langle x_{i} \rangle ^2$. 
To first order in $w$, Eq.~\eqref{Delta_X_Final} is a multi-level Price equation \cite{Price_1972,Price_Unity,Jeler_2018,Gardner_2020,Okasha_2006} (see Supplement Section \ref{1.2.4} for details), which involves a tradeoff between two terms:
\begin{itemize}
    \item $\Delta X_a (\tau)\geq 0$ corresponds to between-deme selection. Cooperators bring a collective benefit to a deme's growth, and demes that start with more cooperators grow faster, thereby yielding a larger contribution to the overall cooperator fraction. The greater the heterogeneity in the initial deme compositions, the stronger this effect is. 
    \item $-\Delta X_b (\tau)$ corresponds to within-deme selection. It satisfies $-\Delta X_b (\tau)\leq 0$, see Supplement Section \ref{1.2.2}. Qualitatively, this is because cooperators are disadvantaged within each deme due to the cost of cooperation, and have a lower fitness than cheaters (see Model and Methods), yielding a decrease of cooperator fraction in each deme.
\end{itemize}
The overall cooperator fraction grows if $\Delta X(\tau)>0$, i.e.\ if the global benefit of cooperation is greater than the local cost, that is $\Delta X_a(\tau)>\Delta X_b(\tau)$. 

\begin{figure}[H]
    \centering
    \includegraphics[width=0.6\textwidth]{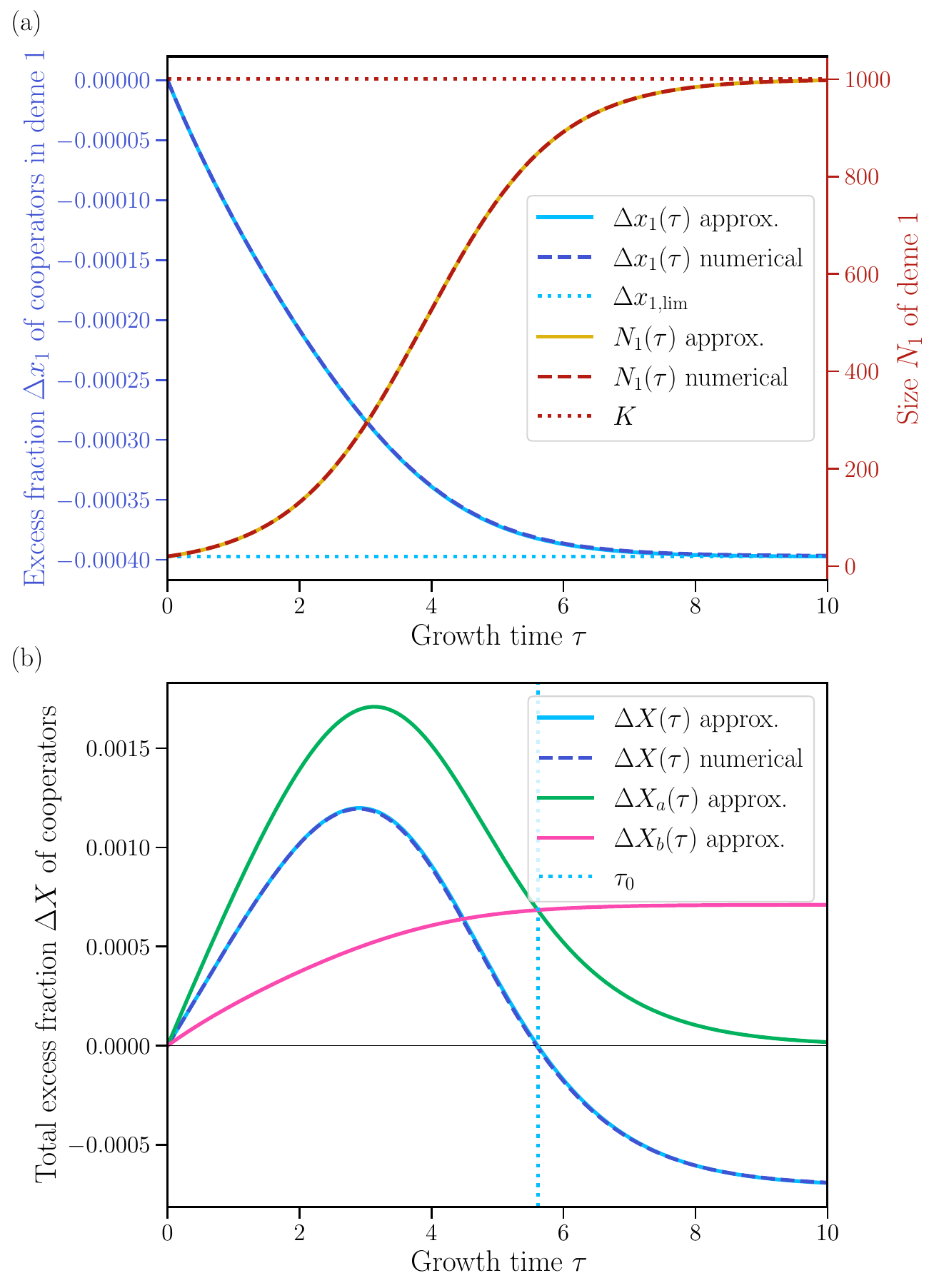}
    \caption{\textbf{Growth phase: evolution of the fraction of cooperators in a deme and in the population.} \textbf{(a)} Excess fraction $\Delta x_1(\tau) = x_1(\tau) - x_1(0)$ of cooperators in one deme, namely deme $1$, and size $N_1$ of this deme, versus growth time $\tau$. We show the numerical solution of Eq.~\eqref{equation_diff} and the analytical approximation of Eq.~\eqref{approximation}, obtained in the weak selection regime. We indicate the long-time asymptotic value $\Delta x_{1,\text{lim}}$ of $\Delta x_1$, obtained from Eq.~\eqref{approximation}, and the carrying capacity $K$ (horizontal dotted lines).
    \textbf{(b)} Excess fraction $\Delta X (\tau)= X(\tau) - X(0)$ of cooperators in the total population versus growth time $\tau$. As in (a), we show the numerical solution and the analytical approximation, here given by Eq.~\eqref{Delta_X_Final}. In addition, we show the two terms from Eq.~\eqref{Delta_Xa} and Eq.~\eqref{Delta_Xb} into which $\Delta X(\tau)$ can be decomposed.
    We also show the largest time $\tau_0>0$ such that $\Delta X (\tau) \ge 0$  (vertical dotted line). In both panels, results are obtained for a population of $D=5$ demes, each with initial size $B=20$ and carrying capacity $K = 1000$, using initial compositions $x_1(\tau=0)=1/10$, $x_2(0)=1/4$, $x_3(0)=1/2$, $x_4(0)=3/4$ and $x_5(0)=9/10$, fitness intensity $w=10^{-3}$, benefit $b=10$ and cost $c=1$.}
    \label{Growth_Phase}
\end{figure}

Fig. \ref{Growth_Phase} shows a situation where $\Delta X(\tau)>0$ transiently, while $x_i(\tau)$ decreases in each deme $i$ (e.g.\ in deme 1, see Fig. \ref{Growth_Phase}(a)). We further observe that our analytical approximations in Eq.~\eqref{approximation} and Eq.~\eqref{Delta_X_Final} agree well with numerical results in the weak selection regime.  

Fig. \ref{Growth_Phase}(b) further shows the time evolution of $\Delta X_a$ and $\Delta X_b$. We observe that the between-group selection term $\Delta X_a$ transiently exceeds the within-group term $\Delta X_b$, leading to $\Delta X(\tau)>0$. Later, $\Delta X_a$ decreases and tends to zero as deme size tends to their carrying capacity $K$ (see Fig. \ref{Growth_Phase}(a)), and the size contrast between demes vanishes. We then have $\Delta X(\tau)<0$. 
In this model, if there is a sufficient heterogeneity in the initial cooperator fractions across different demes, and if demes have not reached saturation by the end of the growth phase, then the resulting contrast in deme sizes can yield $\Delta X(\tau)>0$. This situation is known as Simpson's paradox: global cooperator fraction increases even though local cooperator fraction decreases in each deme, i.e. $\Delta x_i(\tau) = x_i(\tau)- x_i(0) <0$ for all $i$. Such an effect was previously described in experiments and in modeling studies \cite{Simpson_2009,Simpson_2010,Becker_2018,Cremer_2010,Cremer_2011,Cremer_2012,Melbinger_2015, Review_2019}. 
In particular, models with cycles of growth, merging and dilution were investigated in Refs. \cite{Cremer_2010,Cremer_2011,Cremer_2012, Melbinger_2015, Review_2019}. 
While the advantage of cooperation was investigated in these previous studies, our analytical resolution of the dynamics in the weak selection regime allows us to analytically predict the transient advantage of cooperation 
and to quantitatively interpret it as a tradeoff between intra-deme and inter-deme dynamics. The transient nature of the benefit of cooperation is due to growth saturation. Indeed, in Supplement Section~\ref{3.1}, we consider the case of continued exponential growth. In that case, when cooperation is promoted during growth at the scale of the full population, the effect increases over time, because the difference between the sizes of demes with a high and a smaller fraction of cooperators keeps increasing. Note that in the presence of growth saturation, if the carrying capacity increases with the cooperator fraction, the benefit of cooperation can increase for longer growth times~\cite{Cremer_2010,Cremer_2011,Cremer_2012,Melbinger_2015,Arenzon23}. Here, for the sake of simplicity, we have assumed that $K$ does not depend on population composition. 

\begin{figure}[H]
    \centering
    \includegraphics[width=0.7\textwidth]{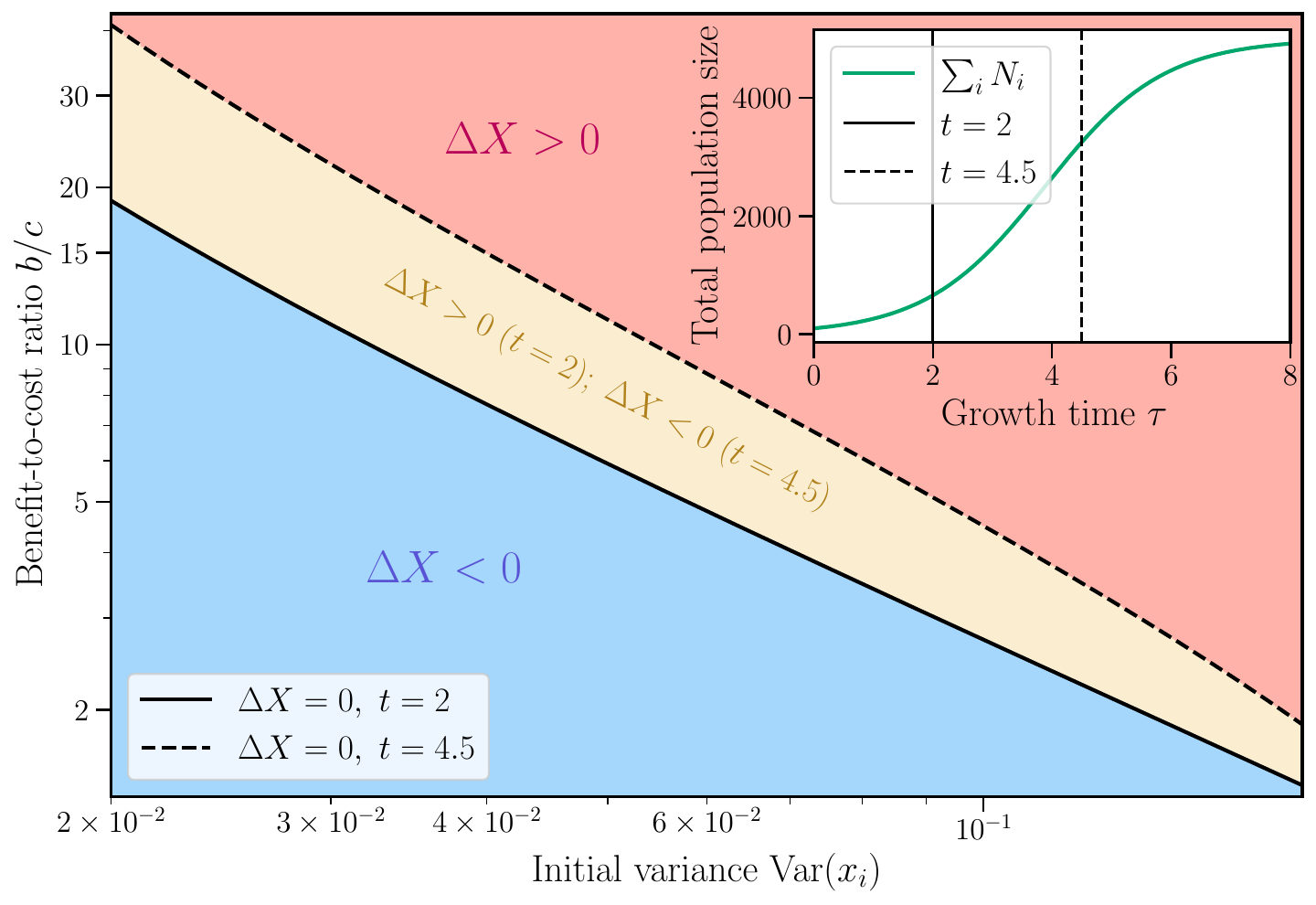}
    \caption{\textbf{Impact of the benefit-to-cost ratio  and of the variance in deme compositions on the evolution of cooperator fraction during growth.} We show a phase diagram, representing whether $\Delta X >0$ or $\Delta X<0$, as a function of the benefit-to-cost ratio $b/c$ and the variance $\mathrm{Var}(x_i)$ of the initial deme compositions,  for two different durations of the growth phase. The separation between zones, namely $\Delta X = 0$, is given by Eq.~\eqref{Condition_Growth} of the Supplement. 
    Results are obtained for a population of $D=5$ demes, each with initial size $B=20$ and carrying capacity $K = 1000$, using different initial compositions with mean $\langle x_i \rangle = 0.5$, fitness intensity $w=10^{-3}$, cost $c=1$ and growth phase durations $t=2$ and $t=4.5$.  The inset shows the analytical approximation for the the total population size $\sum_{i=1}^D N_i$ versus growth time $\tau$, using Eq.~\eqref{approximation} and the same parameter values, for $b = 10$.}
    \label{Phase_Diagram}
\end{figure}

For a given growth time $t$ and for given initial deme compositions, Eq.~\eqref{Delta_X_Final} can be used to compute the threshold value of the benefit-to-cost ratio $b/c$ beyond which cooperation is promoted in the structured population during growth, see Supplement Section \ref{1.2.3}. The smaller the variance in the initial deme compositions, the larger $b/c$ must be for cooperation to be favored during growth, see Fig. \ref{Phase_Diagram}. Assuming that the variance in the initial deme compositions is large enough and the growth time not too short, one can show that this threshold value of $b/c$ increases with time (matching the two cases shown in Fig. \ref{Phase_Diagram}), and that cooperation becomes impossible beyond a certain growth time, see Supplement Section~\ref{1.2.3}. Qualitatively, this is because of the saturation of demes, as the carrying capacity $K$ is here assumed to be independent of cooperator fraction, which entails that cooperator-rich demes lose their advantage of being larger as saturation is reached.

\subsection*{An effective fitness parameter determines if cooperation is favored on average under cycles of dilution-growth-merging}

So far, we determined the conditions that favor cooperation during a given deterministic growth phase. Let us now consider a full cycle of dilution-growth-merging, which involves a stochastic dilution step via binomial sampling, and ask whether cooperation is advantaged on average during such a cycle. For this, we consider the expectation value under the binomial law used for sampling of the difference $\Delta X^\prime = X^\prime_{n}-X^\prime_{n-1}$ in overall cooperator fraction, taken between the end of the growth phase of one cycle (say cycle $n-1$) and the end of the growth phase of the next one ($n$). It can be expressed as (see Supplement Section \ref{1.3} for details)
\begin{align}
    \mathbb{E}\left[\Delta X^\prime\right] = w s(t) X_{n-1}^\prime\left(1-X_{n-1}^\prime\right) + \mathcal{O}\left(w^2\right)  \; , \label{effective_fitness_full}
\end{align}
where $s(t)$ is an effective fitness that depends on time and on model parameters:
\begin{equation}
    s   (t) = \frac{D-1}{DB}\frac{t(b-c)}{1+\frac{B}{K-B}e^t}  -\frac{B-1}{B}\left[t(c-b)+b\log\left(\frac{B(K-1)e^{t}+B-K}{K(B-1)}\right)-c \log\left(\frac{B(e^{t}-1)+K}{K}\right)\right]\,. \label{effective_fitness} 
\end{equation}

The effective fitness $s(t)$ features an interplay between between-deme selection (first term, nonnegative, coming from the expectation of $\Delta X_a$ in Eq.~\eqref{Delta_X_Final}) and within-deme selection (second term, nonpositive, coming from the expectation of $-\Delta X_b$ in Eq.~\eqref{Delta_X_Final}). Eq.~\eqref{effective_fitness_full} shows that, 
assuming $0<X'_{n-1}<1$, cooperation is favored on average over one cycle of dilution-growth-merging (i.e.\ $\mathbb{E}\left[\Delta X^\prime\right] > 0$) if and only if
\begin{equation}
    s(t) > 0\,. \label{Condition_Full_Cycle}
\end{equation}
For brevity, we will henceforth omit the explicit time dependence of $s$. 
For $s$ to be positive, the dilution step needs to generate enough variance in the initial deme compositions to yield a Simpson's paradox scenario during the subsequent growth phase. For this, it is important that the bottleneck size $B$ is small enough, as the expectation value of the variance of deme compositions just after dilution is inversely proportional to $B$ (see Supplement Section \ref{1.3}). While this point was previously highlighted \cite{Simpson_2009,Simpson_2010,Cremer_2010,Cremer_2011,Cremer_2012,Melbinger_2015,Review_2019}, here, we provide an analytical condition for cooperation to be favored on average under cycles of dilution-growth-merging. Note that it can be expressed as a threshold value of the benefit-to-cost ratio $b/c$ beyond which cooperation is favored, see Supplement Section~\ref{1.3.1}. Studying $s$ also allows to investigate extreme parameter regimes, see Supplement Section~\ref{extrreg}. In particular, for large growth times, all demes saturate at the carrying capacity and cooperation cannot be promoted. Conversely, in the limit of large carrying capacities, the promotion of cooperation is not transient but increases, as exponential growth continues. Finally, for large bottleneck sizes, cooperation cannot be promoted, because small bottlenecks are needed to obtain contrast across demes after the dilution step.

While we derived Eq.~\eqref{Condition_Full_Cycle} by focusing on one cycle, this condition does not depend on the precise cycle considered (e.g.\ on the initial cooperator proportions in each deme at the beginning of it). Hence, it is a general condition, valid for any cycle throughout the dynamics and for any number of cycles. In our model, cooperation can be favored in a clique-structured population with weak selection, under repeated cycles of dilution, logistic growth and merging, if and only if Eq.~\eqref{Condition_Full_Cycle} is satisfied. Hard selection, implemented here by merging and dilution, plays a fundamental role, as it allows to promote the collective benefit of cooperation. Indeed, it is via this mechanism that larger demes contribute more to the next bottleneck state. Furthermore, the benefit of cooperation is maintained across cycles through stringent dilution events that yield large compositional variance between demes at each bottleneck. 

\subsection*{Analytical calculation of the fixation probability of cooperator mutants under weak selection}

If cooperation appears through a mutation, it will first be present in a single individual, whose lineage will thus initially be subject to strong finite-size fluctuations (known as genetic drift). In the present model, where growth is deterministic, stochasticity arises from the dilution step, which is modeled by binomial sampling. What is the probability that cooperators ultimately take over the whole population, starting from any initial population composition, e.g.\ from a single cooperator mutant? To address this question, we need to go beyond averaged dynamics, and to perform a stochastic treatment of the system under successive cycles of growth, merging and dilution. A challenge is that we cannot assume that deme sizes are large, given that small bottlenecks and strong variance are necessary for cooperation to be favored on average. This precludes the use of the branching process approximation~\cite{haldane_amathematical_1927,harris_thetheory_1963,Abbara_2023} or of the diffusion approximation~\cite{kimura_diff64} at the scale of demes, since they both require $B\gg 1$ to hold. 
Hence, we perform a calculation of the fixation probability in the weak selection regime $w\ll 1$, without making any assumption on $B$.  

Let us denote by $i_k$ the initial number of cooperators in each deme $k\in\{1,\dots,D\}$ at a bottleneck, and introduce $\bm{i}=(i_1,\dots,i_D)$. Our goal is to determine the fixation probability $p_{\bm{i}}$ of cooperators starting from any initial population composition $\bm{i}$. We have $p_{(0,\dots,0)}=0$ and $p_{(B,\dots,B)}=1$. Let $P_{\bm{i} \rightarrow \bm{j}}$ be the probability of sampling $\bm{j}=(j_1,\dots,j_D)$ cooperators to form each deme's new bottleneck upon a dilution step, after growth and merging, if the previous bottleneck had $\bm{i}=(i_1,\dots,i_D)$ cooperators. Since the dilution step is assumed to proceed through independent binomial samplings (i.e.\ with replacement),  
we have
\begin{equation}
    P_{\bm{i} \rightarrow \bm{j}}=\prod_{k=1}^D \binom{B}{j_k}\left(X^\prime\right)^{j_k}\left(1-X^\prime\right)^{B-j_k} \label{transition_proba} \; ,
\end{equation}
where $X^\prime = X + \Delta X$ denotes the overall cooperator fraction at the end of growth, with $X= \sum_{k=1}^D i_k /(DB)$ the overall initial cooperator fraction before growth and $\Delta X = \Delta X(t)$ given by Eq.~\eqref{Delta_X_Final} with $\tau=t$. For this Markov chain, the fixation probability $p_{\bm{i}}$ is the probability of eventually reaching the absorbing state $(B,\dots,B)$ from the initial configuration $\bm{i}$. Considering all possible states after the first step of the Markov chain, we can write a recurrence relation on $p_{\bm{i}}$ (which is the discrete analog of a steady-state Kolmogorov backward equation):
\begin{equation}
p_{\bm{i}}=\sum_{j_1=0,...,j_D=0}^B P_{\bm{i} \rightarrow \bm{j}} p_{\bm{j}} \; . \label{backward}
\end{equation}
Performing a first-order expansion of Eq.~\eqref{backward} in the weak selection regime 
(See Supplement Section \ref{1.4} for details), we obtain the following expression for the fixation probability:
\begin{align}
    p_{\bm{i}} =  X + \Delta X + wsDBX(1-X) + \mathcal{O}\left(w^2\right) \label{fixation_probability} \; ,
\end{align}
where $\Delta X = \Delta X(t)=\mathcal{O}(w)$ is given by Eq.~\eqref{Delta_X_Final} and characterizes the evolution of the cooperator fraction during the first growth phase, while $s$ is the effective fitness parameter defined in Eq.~\eqref{effective_fitness} and characterizes the cycles of dilution-growth-merging. Note that the fixation probability of defectors, written in terms of the fraction of defectors $Y=1-X$, takes the same form as the one of cooperators in Eq.~\eqref{fixation_probability}, but with the opposite effective fitness coefficient, see Supplement Section~\ref{1.4.1true}. 
Eq.~\eqref{fixation_probability} is an analytical formula that holds for any initial population composition $\bm{i}$ in a spatially structured population undergoing cycles of logistic growth, merging and dilution, in the weak selection regime. It shows that whether cooperation is more or less likely to take over depends on the first growth phase, via $\Delta X$, and on the subsequent cycles of dilution-growth-merging, via $s$, in addition to the initial composition, via $X$. 
Importantly, the result in Eq.~\eqref{fixation_probability} holds for small bottleneck sizes, contrary to results obtained using branching process or diffusion approximations. Note that our weak selection assumption requires $wDB\ll1$, which entails that the domain of validity of Eq.~\eqref{fixation_probability} is different from that of a branching process approximation (which, for fitness intensity $w$ and population size $DB$, assumes $1/(DB)\ll w\ll 1$ and $DB\gg 1$) or of a diffusion approximation (which assumes $w\sim 1/(DB)$ and $DB\gg 1$).

\subsection*{Cooperator fixation can be favored by spatial structure with merging dynamics}

Let us consider as reference neutral mutants, i.e.\ mutants with no fitness difference from the wild-type at any frequency, and let us compare cooperators to this baseline. The fixation probability of cooperators is greater than that of neutral mutants starting from the same initial proportions if $p_{\bm{i}} > X$. Using Eq.~\eqref{fixation_probability}, this condition can be expressed as $\Delta X + wsDBX(1-X) >0$. It allows us to predict the threshold value of the benefit-to-cost ratio $b/c$ beyond which cooperator fixation is promoted in our spatially structured population, see Supplement Section \ref{1.4.1}. 

Fig. \ref{Fixation_Probability_Deterministic} shows a comparison between our analytical expression of the fixation probability in Eq.~\eqref{fixation_probability} and numerical simulation results, starting from a single cooperator mutant. For this, we performed exact stochastic simulations of our model (with cycles of deterministic growth, merging and dilution via binomial sampling), and estimated $p_{\bm{i}}$ as the proportion of simulations (all starting from $\bm{i}$) that reach the absorbing state $X=1$ (as opposed to the other absorbing state $X=0$). We obtain a very good agreement between analytical and numerical results. We observe that, as the benefit-to-cost ratio $b/c$ increases, the fixation probability of a cooperator increases as well. When $b/c$ reaches the predicted threshold value, the fixation probability of a cooperator becomes larger than that of a neutral mutant. This stands in strong contrast with a well-mixed population of the same size (also shown in Fig. \ref{Fixation_Probability_Deterministic}), where a cooperator is never more likely to fix than a neutral mutant, because of the cost of cooperation. Here, cooperation is promoted by the spatial structure considered, and our analytical calculations allow us to determine when and how much this happens, provided that selection is weak. 

\begin{figure}[H]
    \centering
    \includegraphics[width=0.7\textwidth]{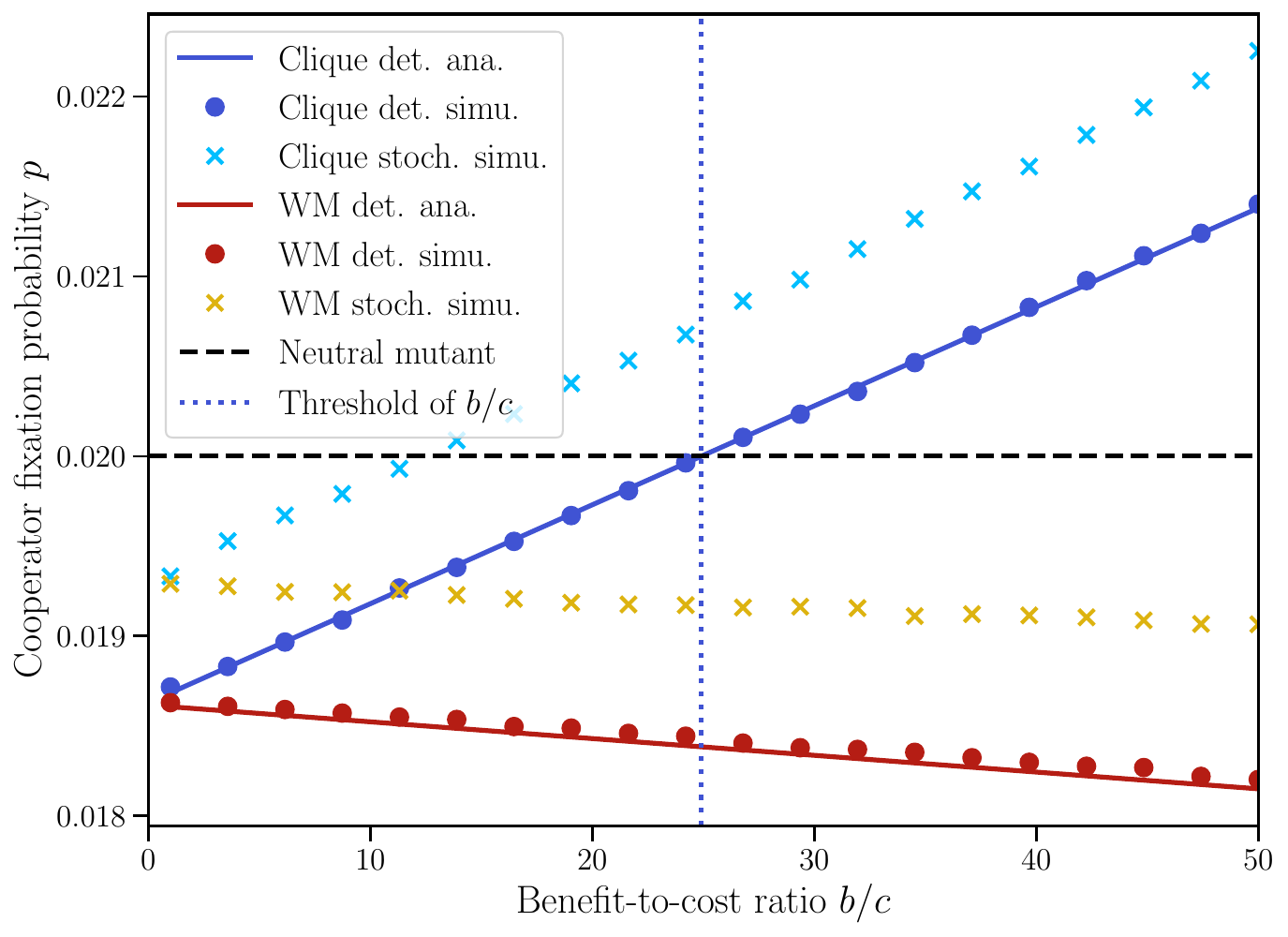}
    \caption{\textbf{Fixation of cooperation.} We show the fixation probability $p$ of cooperation, starting from a single cooperator in the whole population, versus the benefit-to-cost ratio $b/c$, both for our structured population and for a well-mixed (``WM'') population with the same total bottleneck size $DB$ and carrying capacity $DK$. Round markers: numerical simulation (``simu.'') results with deterministic growth (``det.''); solid lines: corresponding analytical predictions (``ana.'') from Eq.~\eqref{fixation_probability}. The threshold of $b/c$ beyond which cooperator fixation is predicted to be promoted (with deterministic growth), given by Eq.~\eqref{Condition_Finale}, is shown as a vertical dotted blue line. Cross markers: numerical simulation results with stochastic growth (``stoch.''). The fixation probability of a neutral mutant, $p=1/(DB)$, is shown for reference (horizontal dashed line). Parameter values: $D=5$, $B=10$, $K=1000$ for a structured population; $D=1$, $B=50$, $K=5000$ for a well-mixed population; for both, $t=3$, $w=5 \times 10^{-4}$ and $c=1$. Simulation results are obtained over $N=10^8$ realizations.}
    \label{Fixation_Probability_Deterministic}
\end{figure}

\subsection*{Stochastic growth further enhances the promotion of cooperation}

In our analysis, we have assumed deterministic logistic growth. We showed that for cooperation to be maintained, the bottleneck size $B$ needs to be small enough for the dilution step to maintain a high variance in deme compositions. Given these small bottleneck sizes, stochastic effects associated with demographic fluctuations during the growth phase can in fact have an important impact. How do our conclusions change if the growth phase is stochastic? To address this question, we consider that each individual can divide with a given rate, matching the one used in our deterministic growth model. We perform simulations of this stochastic growth phase using a Gillespie algorithm, see Supplement Section \ref{2} for details. The merging and dilution steps are unchanged. Note that this modified model comprises two sources of stochasticity, namely the dilution process and the growth process.

In Fig.~\ref{Fixation_Probability_Deterministic}, we compare the fixation probability results in simulations using deterministic and stochastic growth. We find that the fixation probability of a cooperator mutant is higher with stochastic growth than with deterministic growth. 
Hence, demographic fluctuations during growth can further enhance the promotion of cooperation. As previously noted in Ref.~\cite{Cremer_2010}, this is because fluctuations are amplified asymmetrically in this model, in the sense that growth is enhanced if an additional cooperator appears due to a division, but hindered if an additional cheater appears. Our simulation results further suggest that the analytical conditions we derived with deterministic growth provide sufficient conditions for cooperation to be promoted, even with stochastic growth.

\section*{Discussion}

In this work, we considered a minimal model for the evolution of cooperation in deme-structured populations under hard selection. Cooperative interactions were modeled by the prisoner's dilemma, and hard selection by dilution-growth-merging cycles, where each deme contributes to the next bottleneck state in proportion to its size after a phase of local growth. We initially assumed deterministic local growth, but we described dilution via stochastic sampling, and we assumed weak selection. We analytically determined the time evolution of the cooperator fraction during logistic growth, and interpreted this result as a multi-level Price equation, highlighting a tradeoff between within-deme dynamics, where cooperators are disadvantaged, and between-deme dynamics, where demes with more cooperators are advantaged. We then analyzed dilution-growth-merging cycles and derived an analytical condition for the cooperator fraction to increase on average under that dynamics. This requires small bottleneck deme sizes, which generate sufficient variance in deme composition for cooperation to be favored in the whole structured population. Next, we analytically expressed the fixation probability of cooperators under weak selection, without assuming large demes. This yielded a criterion for cooperators to have a higher fixation probability than neutral mutants, and hence for cooperation to be advantaged by natural selection in the spatially structured population. Finally, numerical simulations showed that stochastic growth further promotes cooperation.

A key insight is that, in our simple model of cooperation in deme-structured populations, the mode of selection is decisive: while cooperation cannot be favored under soft selection~\cite{Moawad_2023}, the present study shows that it can be favored under hard selection. Here, we implemented 
hard selection through the dilution-growth-merging process~\cite{Cremer_2012,Melbinger_2015}, whereby demes contribute to the next bottleneck state in proportion to their sizes after growth. As a result, demes with a higher fraction of cooperators can contribute more to the next bottleneck, thanks to their growth advantage. This growth advantage is transient under logistic growth but permanent under continued exponential growth. This coupling between the dynamics of population size and composition has been highlighted previously in Refs.~\cite{Simpson_2009,Simpson_2010,Becker_2018,Cremer_2010,Cremer_2011,Cremer_2012,Melbinger_2015}. Our study connects these results with those of models on graphs with a single individual per node, where spatial structure can favor cooperation under certain individual-level update rules ~\cite{Nowak_2006,ohtsuki2006evolutionary}.

Our work opens several perspectives for further theoretical analysis. First, for simplicity, we assumed that the carrying capacity of a deme is independent of its cooperator fraction. However, a higher cooperator fraction can be associated to a larger carrying capacity~\cite{Simpson_2009,Cremer_2010,Cremer_2011,Cremer_2012,Melbinger_2015}, which enhances the transient advantage of cooperation during logistic growth, and introduces an additional mechanism by which spatial structure can promote cooperation, namely group fixation: demes where cooperators have fixed then become larger than other demes~\cite{Cremer_2012}. Extending our analytical calculations to this case would therefore be of interest. Second, we assumed deterministic growth in most of our analysis, but we observed numerically that stochastic growth can enhance the fixation of cooperators. Further analytical insight into this aspect could be obtained using a van Kampen expansion, as in Refs.~\cite{Cremer_2010,Cremer_2011,Cremer_2012,Melbinger_2015}. A challenge, however, is that this approach relies on an expansion in large system sizes, whereas the promotion of cooperation necessitates small bottleneck sizes. Moreover, while valuable insight was obtained in Refs.~\cite{Cremer_2010,Cremer_2011,Cremer_2012,Melbinger_2015} by focusing on the early stages of growth, analytically determining conditions for cooperation to be promoted would likely require considering the full growth phase. More generally, stochastic growth could be incorporated using stochastic differential equations, but existing work generally focuses on the large-population regime \cite{Bhat2024.02.19.580940}. Third, our analytical expression of the fixation probability assumes weak selection but remains valid for small bottleneck sizes. Thus, a promising direction for future work is to extend our analysis beyond the weak selection regime. Besides, the domain of validity of our current approach differs from that of diffusion and branching process approximations, which assume large deme sizes, and more general descriptions encompassing these different regimes would be highly interesting, in dilution-growth models and in Wright-Fisher, which are closely related~\cite{Abbara_2023}, as well as in models with partial updates, which bridge Wright-Fisher and Moran models \cite{ALEXANDRE2025112030}. 

More generally, while the merging process facilitates analytical progress thanks to its strong symmetries, it would be interesting to study other forms of hard selection in more complex population structures. First, one could consider a clique where each deme contributes in proportion to its productivity (i.e.\ its size after growth), but where the probability that an organism migrates to another deme differs from the probability that it stays in its deme of origin (thus going beyond the specific case of merging). Next, one could consider more  
general graphs, in order to establish a more complete link with results from evolutionary graph theory~\cite{ohtsuki2006evolutionary,Nowak_2006}. Studying other forms of cooperative interactions is also a promising direction and would allow closer links to specific natural or experimental systems. This includes considering games other than the prisoner's dilemma, such as the snowdrift game~\cite{Gore09}, modeling cooperative antibiotic resistance~\cite{hernandez_coupled_2023,hernandez_ecoevolutionary_2024,hernandez-navarro_slow_2024,denk-lobnig_spatial_2025}, and explicitly describing the dynamics of diffusible public goods~\cite{Borenstein13,allen_spatial_2013,menon_public_2015,matsuzawa_social_2016} and resource limitation \cite{waite2015defectors}.  

\section*{Data availability and code availability}
Our code is  freely available at \url{https://github.com/Bitbol-Lab/Cooperation_Merging}.
 
\section*{Acknowledgments}
This project has received funding from the European Research Council (ERC) under the European Union’s Horizon 2020 research and innovation programme (grant agreement No.~851173, to A.-F.~B.).

\newpage

\begin{center}
 {\LARGE \bf Supplementary material}   
\end{center}

\renewcommand{\thefigure}{S\arabic{figure}}
\setcounter{figure}{0}
\renewcommand{\thetable}{S\arabic{table}}
\setcounter{table}{0} 

\tableofcontents


\section{Derivations of analytical expressions}\label{1}

\subsection{Intra-deme growth dynamics}\label{1.1}

To find an approximate analytical solution of Eq.~\eqref{equation_diff} in the weak selection regime, we perform a first-order perturbative expansion in the fitness parameter $w$:
\begin{align}
    N(\tau) &= N_0(\tau) + w N_1(\tau) + \mathcal{O}(w^2)\,, \nonumber \\
    x(\tau) &= x_0(\tau) + w x_1(\tau) + \mathcal{O}(w^2) \,.
\end{align}
Substituting this into Eq.~\eqref{equation_diff} and identifying terms at each order leads to
\begin{align}
    &\frac{dN_0}{d\tau} = N_0\left(1-\frac{N_0}{K}\right)\,, \label{N_0} \\
    &\frac{dN_1}{d\tau} = [(b-c)x_0-1]N_0\left(1-\frac{N_0}{K}\right) + N_1\left(1-\frac{2N_0}{K}\right)\,, \label{N_1} \\
    &\frac{dx_0}{d\tau} = 0\,, \label{x_0} \\
    &\frac{dx_1}{d\tau} = -x_0(1-x_0)\left(\frac{b}{N_0-1}+c\right)\left(1-\frac{N_0}{K}\right)\,. \label{x_1}
\end{align}
Eq.~\eqref{N_0} is a logistic equation with solution $N_0(\tau)=K /\left[1+(K-B)e^{-\tau}/B\right]$, where we used the initial condition $N_0(0)=B$. Eq.~\eqref{x_0} leads to $x_0 (\tau)= x_0(0)\equiv x_0$. Using these expressions then allows us to solve Eqs.~\eqref{N_1} and \eqref{x_1}. This yields the explicit analytical expressions in Eq.~\eqref{approximation}, i.e.
\begin{align}
    N(\tau) &= 
    \frac{K}{1+\frac{K-B}{B}e^{-\tau}} \left(1 + w  \frac{\tau[(b-c)x_0-1]}{1+\frac{B}{K-B}e^\tau}\right) + \mathcal{O}(w^2) \,, \label{sol_N} \\
    x(\tau) &= x_0-w x_0(1-x_0)\left[\tau(c-b)+b\log\left(\frac{B(K-1)e^{\tau}+B-K}{K(B-1)}\right)-c\log\left(\frac{B(e^{\tau}-1)+K}{K}\right)\right] \nonumber \\
    & + \mathcal{O}(w^2) \,.\label{sol_x} 
\end{align}

\subsection{Full-graph growth dynamics}\label{1.2}

\subsubsection{Dynamics of the overall fraction of cooperators during growth}\label{1.2.1}

Now that we have described the  dynamics of growth within each well-mixed deme, let us turn to the whole graph with $D$ demes indexed by $i\in\{ 1,\dots, D\}$, each of them starting from the same initial bottleneck size $B$. Let us denote by $x_i = x_i(0)$ the initial fraction of cooperators in deme $i$, and by $N_i^\prime = N_i(\tau)$ (resp.\ $x^\prime_i = x_i(\tau)$) the size (resp.\ fraction of cooperators) of deme $i$ at time $\tau$ during the growth phase. 
Finally, let us introduce the initial and current global fractions of cooperators in the whole spatially structured population as:
\begin{equation}
X= \frac{1}{D}\sum_{i=1}^D x_i= \langle x_i \rangle \quad \text{and} \quad X^\prime= \frac{\sum_{i=1}^D N_i^\prime x^\prime_i}{\sum_{i=1}^D N_i^\prime}\,, \label{xX}
\end{equation}
respectively. Here, the notation $\langle . \rangle$ denotes an average over demes, or equivalently, over the whole population at a bottleneck, since all demes have the same bottleneck size $B$. 

Let us now perform a first-order expansion in $w \ll 1$ to obtain an explicit analytical expression of $X^\prime$ in the weak selection regime:
\begin{equation}
X^\prime = X^\prime(w=0) + w  \frac{\partial X^\prime}{\partial w} (w=0) + \mathcal{O}\left(w^2\right)\,.
\label{taylor}
\end{equation}
Eq.~\eqref{sol_x} gives $X^\prime (w=0) = X$. Then, we write
\begin{align}
    \frac{\partial X^\prime}{\partial w} &= \frac{\sum_i \left(\frac{\partial N_i^\prime }{\partial w} x_i^\prime + N_i^\prime \frac{\partial x_i^\prime}{\partial w}\right)\sum_j N_j^\prime - \sum_iN_i^\prime x_i^\prime \sum_j \frac{\partial N_j^\prime }{\partial w}}{\left(\sum_i N_i^\prime \right)^2} \,.
    \label{piece1}
\end{align}
Next, we express the different terms involved in Eq.~\eqref{piece1}, for $w=0$, using Eq.~\eqref{sol_N} and Eq.~\eqref{sol_x}. We obtain
\begin{align}
\frac{\partial N_i^\prime}{\partial w} (w=0) &= \frac{K}{1+\frac{K-B}{B}e^{-\tau}}\, \frac{\tau[(b-c)x_i -1]}{1+\frac{B}{K-B}e^\tau}\,,
\label{piece2}\\
\frac{\partial x_i^\prime}{\partial w} (w=0) &= -x_i(1-x_i)\left[\tau(c-b)+b\log\left(\frac{B(K-1)e^{\tau}+B-K}{K(B-1)}\right)-c\log\left(\frac{B(e^{\tau}-1)+K}{K}\right)\right]\,.
\label{piece3}
\end{align}
Combining Eqs.~\eqref{piece1}, \eqref{piece2} and \eqref{piece3}, we obtain
\begin{align}
    \frac{\partial X^\prime}{\partial w} (w=0) &= - \left \langle x_i(1-x_i)\right \rangle \left[\tau(c-b)+b\log\left(\frac{B(K-1)e^{\tau}+B-K}{K(B-1)}\right)-c \log\left(\frac{B(e^{\tau}-1)+K}{K}\right)\right] \nonumber \\
    & + \mathrm{Var}(x_i)\frac{\tau  (b-c)}{1+\frac{B}{K-B}e^{\tau}} \; ,
    \label{dXprime}
\end{align}
where $\mathrm{Var}(x_i)= \left \langle x_i^2 \right \rangle - \left \langle x_i \right \rangle ^2$ denotes the variance of the initial group compositions. Combining Eq.~\eqref{dXprime} and Eq.~\eqref{taylor}, and using $X^\prime (w=0) = X$, yields Eq.~\eqref{Delta_X_Final} in the main text, i.e.
\begin{align}
    \Delta X(\tau) &=  X(\tau)- X(0)=w\left[\Delta X_a (\tau) - \Delta X_b (\tau)\right] + \mathcal{O}\left(w^2\right) \label{Delta_X_Final_SI} ,\;\;\textrm{with}\\
    \Delta X_a (\tau) &= \mathrm{Var}(x_i) \frac{\tau  (b-c)}{1+\frac{B}{K-B}e^\tau}\; ,\;\;\textrm{and}\\
    \Delta X_b(\tau) &= \left \langle x_i(1-x_i)\right \rangle \left[\tau(c-b)+b\log\left(\frac{B(K-1)e^{\tau}+B-K}{K(B-1)}\right)-c \log\left(\frac{B(e^{\tau}-1)+K}{K}\right)\right] \label{DeltaXb}\; ,
\end{align}

\subsubsection{Proof that $\Delta X_b$ is nonnegative} \label{1.2.2}

Here we show that $\Delta X_b(\tau)$, defined in Eq.~\eqref{DeltaXb}, is nonnegative. First, we notice that $\Delta X_b (0)=0$. 
Second, for any time $\tau \in [0,t]$ during the growth phase, we have
\begin{align}
    \frac{\partial \Delta X_b}{\partial \tau} (\tau)
    & = \langle x_i(1-x_i )\rangle \left[ b \left(\frac{B(K-1)}{B(K-1)-(K-B)e^{-\tau}}-1\right) + c \left(1-\frac{B}{(K-B)e^{-\tau}+B}\right) \right]\,.
\end{align}
Since $K>B>1$, $b>0$ and $c>0$, this yields $\forall \tau \in [0,t], \,\,\partial \Delta X_b / \partial \tau \ge 0$. Consequently, we obtain $\forall \tau \in [0,t],\,\,\Delta X_b (\tau) \ge 0$.  

\subsubsection{Condition on the benefit-to-cost ratio for cooperation to be favored during growth}\label{1.2.3}

Cooperation is favored by growth if at the end of a growth phase $\tau=t$, we have $\Delta X(t) >0$, where $\Delta X(t)$ is given by Eq.~\eqref{Delta_X_Final_SI} in the weak selection approximation. The threshold value of the benefit-to-cost ratio $b/c$ beyond which cooperation is favored is thus obtained by solving $\Delta X(t) = 0$, which yields:
\begin{equation}
    \frac{b}{c} = \frac{\langle x_i(1-x_i) \rangle \left[t-\log\left(\frac{B(e^{t}-1)+K}{K}\right)\right] + \mathrm{Var}(x_i)\,\frac{t}{1+\frac{B}{K-B}e^t} }{\left \langle x_i(1-x_i)\right \rangle \left[t-\log\left(\frac{B(K-1)e^{t}+B-K}{K(B-1)}\right)\right] + \mathrm{Var}(x_i)\,\frac{t}{1+\frac{B}{K-B}e^t} }\,. \label{Condition_Growth} 
\end{equation}
Eq.~\eqref{Condition_Growth} is used to plot the curves $\Delta X = 0$ in Fig.~\ref{Phase_Diagram} in the main text.

\paragraph{Dependence on growth time.} Let us now investigate how this threshold value of $b/c$ depends on growth time $t$. 
Eq.~\eqref{Condition_Growth} can be written as:
\begin{equation}
    \frac{b}{c} = \theta(t) = \frac{\theta_1(t)}{\theta_2(t)},
\end{equation}
where $\theta_1$ (resp.\ $\theta_2$) represents the numerator (resp.\ denominator) in Eq.~\eqref{Condition_Growth}. We notice that $\forall t \geq 0,  \,\,\theta_1(t)\geq0$, given that $B<K$. Next, we write:
\begin{equation}
    \theta_2(t) = \langle x_i(1-x_i) \rangle f(t) + \mathrm{Var}(x_i) g(t) ,
\end{equation}
and we notice that $f(t) \le 0$ and that $f$ decreases with time for $t \ge 0$. Meanwhile, $g(t) \ge 0$ for $t \ge 0$ and $g$ increases with $t$ for $t<t_1$ but decreases with $t$ for $t>t_1$, with:
\begin{equation}
    t_1 = 1 + W \left(\frac{K-B}{eB}\right),
\end{equation}
where $W$ denotes the Lambert $W$ function. Furthermore, $g (t)\to 0$ as $t\to\infty$. Hence, $\theta_2$ decreases with $t$ for $t > t_1$, and $\theta_2(t)<0$ if $t$ is large enough. 

Let us assume that $\theta_2(t_1) > 0$, which is the case if the variance of cooperator fraction across demes is large enough, specifically if
\begin{equation}
    \frac{\mathrm{Var}(x_i)}{\langle x_i(1-x_i) \rangle} > -\frac{ f(t_1)}{g(t_1)}\,. \label{varcdn}
\end{equation}
Then, there exists $t_*>t_1$ such that $\theta_2(t_*) = 0$, 
and we have $\theta(t) \to \infty$ as $t \to t_*$. 
Let us now show that $\theta^{\prime}(t)>0$ for $t \in [t_1,t_*)$. For this, we write:
\begin{equation}
    \theta(t) = 1 + \langle x_i(1-x_i)\rangle \frac{h(t)}{\theta_2(t)} \; ,
\end{equation}
with
\begin{equation}
    h(t) = \log\left[\frac{B(K-1)e^t + B-K}{\left(B-1\right)\left[B\left(e^t-1\right)+K\right]}\right] \; .
\end{equation}
We have:
\begin{equation}
    \theta^{\prime}(t) = \langle x_i(1-x_i)\rangle \frac{h^{\prime}(t)\theta_2(t)-h(t)\theta_2^{\prime}(t)}{\theta_2^2(t)} \; . \label{derivative}
\end{equation}
One can show that for $t > 0$, $h(t) > 0$ and $h^{\prime}(t) > 0$, since $K>B$, and assuming $B>2$. 
For $t \in [t_1, t_*)$, both terms on the numerator of Eq.~\eqref{derivative}
are positive, since $h(t), h^{\prime}(t), \theta_2(t) > 0$ and $\theta_2^{\prime}(t) < 0$. Hence, $\theta^{\prime}(t) > 0$ for $t \in [t_1, t_*)$. 

To conclude, assuming that the condition in Eq.~\eqref{varcdn} is satisfied, the threshold value $\theta(t)$ of $b/c$ given by Eq.~\eqref{Condition_Growth} increases with growth time for $t \in [t_1,t_*)$ and diverges as $t \to t_*$. In addition, under the same condition, $\theta(t)<0$ for $t>t_*$, meaning that cooperation cannot be promoted in the limit of long growth times. 

\subsubsection{Multi-level Price equation}\label{1.2.4}

Here, we show that, to first order in $w$, Eq.~\eqref{Delta_X_Final_SI}, i.e.\ Eq.~\eqref{Delta_X_Final} in the main text, is a \textit{multi-level Price equation} \cite{Okasha_2006,Jeler_2018,Price_Unity}, where one compares the overall fraction of cooperators in the metapopulation between the beginning and the end of a growth phase, i.e.\ between $\tau=0$ and $\tau=t$. Let us consider $w_i(t)= N_i(t)/B$, which quantifies the growth of deme $i$, and is known as its \textit{Wrightian fitness}~\cite{Wu_Bib_Gokhale}. The discrete multi-level Price equation is a mathematical identity that expresses the variation of an overall trait during growth, here that of the overall cooperator fraction $X$, namely $\Delta X(t)=X(t)-X(0)$, as: 
\begin{equation}
    \Delta X(t) = \frac{1}{\langle w_i(t) \rangle}\left[\mathrm{Cov}{(x_i, w_i(t))} + \langle w_i(t) \Delta x_i(t) \rangle \right] \; , \label{Price_Equation} 
\end{equation}
with $\mathrm{Cov}{(x_i, w_i(t))} = \langle x_i w_i(t) \rangle - \langle x_i \rangle \langle w_i(t) \rangle$. The first term is known as \textit{between-group selection} \cite{Gardner_2020,Jeler_2018,Okasha_2006,Price_Unity}, which is nonnegative in our model because cooperators increase a deme's Wrightian fitness. The second term is known as \textit{within-group selection}, for which cooperators are disadvantaged in our model. To first order in $w$, the two terms of Eq.~\eqref{Price_Equation}, expressed using Eq.~\eqref{sol_N} and Eq.~\eqref{sol_x}, exactly correspond to those of Eq.~\eqref{Delta_X_Final_SI}: $\mathrm{Cov}{(x_i, w_i(t))}/\langle w_i(t) \rangle=w \Delta X_a (t)$, and $\langle w_i(t) \Delta x_i \rangle/ \langle w_i(t) \rangle=-w \Delta X_b (t)$.

\subsection{Full-graph dynamics, one cycle of dilution-growth-merging} \label{1.3}

\subsubsection{Effective fitness parameter over one cycle of dilution-growth-merging} \label{efffit}

Consider the spatially structured population at cycle $n-1$  after a growth and merging step, just before a stochastic dilution step. The global fraction of cooperators is then $X_{n-1}^\prime$. For each deme, the dilution event consists in a binomial sampling of $B$ individuals (with replacement) from the merged population, with probability of success $X_{n-1}^\prime$ to pick a cooperator. This yields a new bottleneck state for all demes. 
Let us denote by $x_i$ the fraction of cooperators after dilution in deme $i$ (i.e.\ at the beginning of the next growth phase, which corresponds to cycle $n$). From the properties of the binomial distribution, we have, $\forall i\in\{1,\dots, D\}$,
\begin{align}
\mathbb{E}\left[x_i\right]&=X_{n-1}^\prime\,,\label{esp1}\\
\mathbb{E}\left[x_i^2\right]&=X_{n-1}^{\prime 2} + \frac{1}{B}X_{n-1}^\prime(1-X_{n-1}^\prime)\,,\label{esp2}
\end{align}
where $\mathbb{E}[.]$ denotes the expectation value under the binomial law, in contrast with $\langle.\rangle$, which denotes a mean over the $D$ demes of the population. 
Let us denote by $X_{n} = \langle x_i \rangle$ the overall fraction of cooperators in the graph after dilution, just before growth, and by $X_{n}^\prime$ the final overall fraction after growth. 

We are interested in the expectation value of the variation $\Delta X^\prime=X_{n}^\prime -X_{n-1}^\prime$ of the overall cooperator fraction during a full cycle of dilution, growth and merging:
\begin{equation}
    \mathbb{E}\left[\Delta X^\prime\right] = \mathbb{E}\left[X_{n}^\prime\right] -X_{n-1}^\prime  = \mathbb{E}\left[X_{n}^\prime -X_{n} \right]= \mathbb{E}\left[X_{n}(t) -X_{n}(0) \right]= \mathbb{E}\left[\Delta X(t)\right] \,,\label{expvar}
\end{equation}
where we used $\mathbb{E}\left[X_{n}\right] = \mathbb{E}\left[\left \langle x_i\right\rangle\right] = \left \langle\mathbb{E}\left[ x_i\right]\right\rangle = X_{n-1}^\prime$. Recall that the growth phase lasts from $\tau=0$ to $\tau=t$.  
Using Eq.~\eqref{Delta_X_Final_SI}, i.e.\ Eq.~\eqref{Delta_X_Final} of the main text, for the final fraction of cooperators after growth given the initial conditions before growth, Eq.~\eqref{expvar} gives:
\begin{align}
    \mathbb{E}\left[\Delta X^\prime\right] &= w\,\mathbb{E}\left[\mathrm{Var}(x_i)\right] \frac{t (b-c) }{1+\frac{B}{K-B}e^t}\nonumber\\&- w\,\mathbb{E}\left[\left \langle x_i(1-x_i)\right \rangle \right]\left[t(c-b)+b\log\left(\frac{B(K-1)e^{t}+B-K}{K(B-1)}\right)-c \log\left(\frac{B(e^{t}-1)+K}{K}\right)\right]\nonumber\\&+\mathcal{O}(w^2). \label{esp5}
\end{align}
Eqs.~\eqref{esp1} and~\eqref{esp2} give:
\begin{align}
    \mathbb{E}\left[\left \langle x_i^2\right\rangle\right] & = \left \langle \mathbb{E}\left[x_i^2\right] \right\rangle= X_{n-1}^{\prime 2} + \frac{1}{B} X_{n-1}^\prime(1-X_{n-1}^\prime)\,, \\
    \mathbb{E}\left[\left \langle x_i\right\rangle^2\right] & = \frac{1}{D^2}\sum_{i,j} \mathbb{E}\left[ x_i x_j\right] = \frac{1}{D^2} \sum_{i} \mathbb{E}\left[ x_i^2\right] + \frac{1}{D^2} \sum_{i\ne j} \mathbb{E}\left[ x_i\right] \mathbb{E}\left[ x_j\right] = X_{n-1}^{\prime 2} + \frac{1}{DB}X_{n-1}^\prime (1-X_{n-1}^\prime) \,,
\end{align}
leading to
\begin{align}
    \mathbb{E}\left[\mathrm{Var}(x_i)\right] & = \mathbb{E}\left[\left \langle x_i^2\right\rangle\right] -  \mathbb{E}\left[\left \langle x_i\right\rangle^2\right] = \frac{D-1}{DB}X_{n-1}^\prime(1-X_{n-1}^\prime) \,,\label{esp3}\\
    \mathbb{E}\left[\left \langle x_i(1-x_i)\right \rangle \right]&=\mathbb{E}\left[\left \langle x_i\right \rangle \right]-\mathbb{E}\left[\left \langle x_i^2\right \rangle \right]=\frac{B-1}{B}X'_{n-1}(1-X'_{n-1})\,.\label{esp4}
\end{align}
Using Eqs.~\eqref{esp3} and~\eqref{esp4}, Eq.~\eqref{esp5} becomes
\begin{align}
    \mathbb{E}\left[\Delta X^\prime\right] = w s (t)  X_{n-1}^\prime\left(1-X_{n-1}^\prime\right)   + \mathcal{O}\left(w^2\right)  \, , \label{effective_fitness_full_SI}
\end{align}
where we introduced the effective fitness parameter
\begin{equation}
    s (t)   = \frac{D-1}{DB}\frac{t(b-c)}{1+\frac{B}{K-B}e^t}  -\frac{B-1}{B}\left[t(c-b)+b\log\left(\frac{B(K-1)e^{t}+B-K}{K(B-1)}\right)-c \log\left(\frac{B(e^{t}-1)+K}{K}\right)\right]\,. \label{effective_fitness_SI} 
\end{equation}
As described in the main text, $s$ summarizes the interplay between within-group et between-group selection, in the spirit of the multi-level Price equation. Cooperation is favored on average in the whole population if $s>0$, and disadvantaged if $s<0$. 

\subsubsection{Condition on the benefit-to-cost ratio for cooperation to be favored on average over a cycle of dilution-growth-merging} \label{1.3.1}

Cooperation is favored on average for a full cycle of dilution-growth-merging if after merging, we have $ \mathbb{E}\left[\Delta X^\prime\right] >0$, where $ \mathbb{E}\left[\Delta X^\prime\right]$ is given by Eq.~\eqref{effective_fitness_full_SI} and Eq.~\eqref{effective_fitness_SI} in the weak selection approximation. The threshold value of the benefit-to-cost ratio $b/c$ beyond which cooperation is favored is thus obtained by solving $s = 0$, which leads to:
\begin{equation}
    \frac{b}{c} = \frac{(B-1)\left[t-\log\left(\frac{B(e^{t}-1)+K}{K}\right)\right] + \frac{D-1}{D}\frac{t}{1+\frac{B}{K-B}e^t} }{(B-1) \left[t-\log\left(\frac{B(K-1)e^{t}+B-K}{K(B-1)}\right)\right] + \frac{D-1}{D} \frac{t}{1+\frac{B}{K-B}e^t} }\,. \label{Condition_Cycle} 
\end{equation}

\subsection{Fixation probability of cooperation} \label{1.4}

\subsubsection{Derivation of the fixation probability of cooperation} \label{1.4.1true}

Here, we give more detail on our calculation of the fixation probability $p_{\bm{i}}$ of cooperators, starting from $\bm{i}=(i_1,\dots,i_D)$ cooperators in each deme. Specifically, in the weak-selection regime $w\ll 1$, we solve Eq.~\eqref{backward} in the main text, which reads
\begin{equation}
p_{\bm{i}}=\sum_{j_1=0,...,j_D=0}^B P_{\bm{i} \rightarrow \bm{j}} p_{\bm{j}} \equiv \sum_{\{j_k\}}P_{\bm{i} \rightarrow \bm{j}} p_{\bm{j}}\; , \label{backward_SI}
\end{equation}
with
\begin{equation}
    P_{\bm{i} \rightarrow \bm{j}}=\prod_{k=1}^D \binom{B}{j_k}\left(X^\prime\right)^{j_k}\left(1-X^\prime\right)^{B-j_k} \label{transition_proba_SI} \; ,
\end{equation}
where $X^\prime = X + \Delta X$ denotes the overall cooperator fraction at the end of growth, with $X= \sum_{k=1}^D i_k /(DB)$ the overall initial cooperator fraction before growth and $\Delta X = \Delta X(t)$ given by Eq.~\eqref{Delta_X_Final_SI}. 

We expand the fixation probability as 
\begin{equation}
p_{\bm{i}}= X + wv_{\bm{i}}+\mathcal{O}\left(w^2\right) \; , \label{fixation_probability_expansion_SI}
\end{equation}
where we employed the fact that the fixation probability in the neutral case $w=0$ is $X$, and where $v_{\bm{i}}$ is the first-order term that we aim to determine. For this, we will expand Eq.~\eqref{backward_SI} to first order in $w$, and solve this equation. 
Expanding the transition probability given by Eq.~\eqref{transition_proba_SI}, we have
\begin{align}
    P_{\bm{i} \rightarrow \bm{j}} &= P_{\bm{i} \rightarrow \bm{j}}^{(0)}+w P_{\bm{i} \rightarrow \bm{j}}^{(1)}+\mathcal{O}(w^2) \label{bin_exp}\,,\,\,\,\textrm{with}\\
    P_{\bm{i} \rightarrow \bm{j}}^{(0)}&= \prod_{k=1}^D \binom{B}{j_k}\left(X\right)^{j_k}\left(1-X\right)^{B-j_k} \label{transition_proba_0}\,,\,\,\,\textrm{and}\\
    P_{\bm{i} \rightarrow \bm{j}}^{(1)}
    &=\frac{C P_{\bm{i} \rightarrow \bm{j}}^{(0)}}{X(1-X)}\left(\sum_{l=1}^D j_l -BDX \right)\label{transition_proba_1} \; ,
\end{align}
where we introduced $C$ such that $\Delta X = wC + \mathcal{O}\left(w^2\right)$, as given by Eq.~\eqref{Delta_X_Final_SI}. 
Using Eq.~\eqref{fixation_probability_expansion_SI} and Eq.~\eqref{bin_exp}, we obtain the first-order expansion of Eq.~\eqref{backward_SI} in $w$: 
\begin{align}
    X + wv_{\bm{i}} + \mathcal{O}(w^2)&=\sum_{\{j_k\}}\left[P_{\bm{i} \rightarrow \bm{j}}^{(0)}+w P_{\bm{i} \rightarrow \bm{j}}^{(1)}+\mathcal{O}(w^2)\right]\left[\frac{1}{DB}\sum_{n=1}^D j_n  + wv_{\bm{j}} + \mathcal{O}(w^2)\right]\nonumber\\
    &=\sum_{\{j_k\}}\left\{\frac{1}{DB}\sum_{n=1}^D j_n P_{\bm{i} \rightarrow \bm{j}}^{(0)}+w\left[P_{\bm{i} \rightarrow \bm{j}}^{(0)}v_{\bm{j}}+\frac{1}{DB}\sum_{n=1}^D j_n P_{\bm{i} \rightarrow \bm{j}}^{(1)}\right]\right\} + \mathcal{O}(w^2) \label{backward_expanded}\,.
\end{align}
Denoting by $\mathbb{E}_{(0)}[.]$ the expectation value under the probability distribution introduced in Eq.~\eqref{transition_proba_0}, and using the properties of binomial distributions, we obtain
\begin{equation}
\sum_{\{j_k\}}\frac{1}{DB}\sum_{n=1}^D j_n P_{\bm{i} \rightarrow \bm{j}}^{(0)}=\frac{1}{DB}\,\mathbb{E}_{(0)}\left[ \sum_{n=1}^D j_n \right]=\frac{1}{B}\,\mathbb{E}_{(0)}[ j_n]=X\,,
\end{equation}
and
\begin{align}
\sum_{\{j_k\}}\frac{1}{DB}\sum_{n=1}^D j_n P_{\bm{i} \rightarrow \bm{j}}^{(1)}&=\frac{C}{DBX(1-X)}\left\{\mathbb{E}_{(0)}\left[ \sum_{n=1}^D j_n\sum_{l=1}^D j_l \right]-BDX\mathbb{E}_{(0)}\left[ \sum_{n=1}^D j_n \right]\right\}\nonumber\\
&=\frac{C}{DBX(1-X)}\left\{D(D-1)\left(\mathbb{E}_{(0)}\left[ j_n\right]\right)^2+D\, \mathbb{E}_{(0)}\left[ j_n^2\right]-BD^2X\,\mathbb{E}_{(0)}\left[ j_n \right]\right\}\nonumber\\
&=\frac{C}{DBX(1-X)}\left\{D(D-1)B^2X^2+D(BX-BX^2+B^2X^2)-B^2D^2X^2\right\}\nonumber\\
&=C\,.
\end{align}
Eq.~\eqref{backward_expanded} thus becomes
\begin{equation}
    X+wv_{\bm{i}}+ \mathcal{O}(w^2) =X+w\sum_{\{j_k\}}P_{\bm{i} \rightarrow \bm{j}}^{(0)}v_{\bm{j}}+ wC + \mathcal{O}(w^2) \label{backward_expanded_n}\,,
\end{equation}
which yields
\begin{align}
    v_{\bm{i}}&=\sum_{\{j_k\}} P_{\bm{i} \rightarrow \bm{j}}^{(0)} v_{\bm{j}} + C =\mathbb{E}_{(0)}\left[ v_{\bm{j}} \right] +C\; . \label{new_backward_SI}
\end{align}
Note that the identification of terms in the expansion of Eq.~\eqref{backward_expanded} requires $wDB\ll1$. Thus, our weak selection assumption requires $w\ll1/(DB)$.

As our goal is to determine $v_{\bm{i}}$, we now aim to solve Eq.~\eqref{new_backward_SI}. In Eq.~\eqref{Delta_X_Final_SI}, we observe that $C$ can be expressed as 
\begin{align}
C&= C_1X + C_2X^2 + C_3 \left \langle x_i^2\right \rangle \label{Cdecompo}\,,\,\,\, \textrm{with}\\
    C_1& = - \left[t(c-b)+b\log\left(\frac{B(K-1)e^{t}+B-K}{K(B-1)}\right)-c \log\left(\frac{B(e^{t}-1)+K}{K}\right)\right] \,,\\
    C_2& = - \frac{t(b-c)}{1+\frac{B}{K-B}e^t}   \,,\\
    C_3& = -C_1 - C_2 \,.
\end{align}
We thus look for a solution of Eq.~\eqref{new_backward_SI} that takes the following form:
\begin{align}
    v_{\bm{i}} &= A_1 X + A_2 X^2 + A_3 \left \langle x_i^2\right \rangle = \frac{A_1}{DB} \sum_{l=1}^D i_l  + \frac{A_2}{\left(DB \right)^2} \left(\sum_{l=1}^D i_l\right)^2 + \frac{A_3}{DB^2} \sum_{l=1}^D i_l^2 \; ,\label{form}
\end{align}
where $A_1$, $A_2$ and $A_3$ are constants to be determined (which do not depend on $\bm{i}$). 
With this expression, we have
\begin{align}
\mathbb{E}_{(0)}\left[ v_{\bm{j}} \right]&=\frac{A_1}{B}\,\mathbb{E}_{(0)}\left[ j_l \right]+\frac{A_2}{\left(DB \right)^2}\,\mathbb{E}_{(0)}\left[ \sum_{n=1}^D j_n\sum_{l=1}^D j_l \right]+\frac{A_3}{B^2}\,\mathbb{E}_{(0)}\left[ j_l^2 \right]\nonumber\\
&=\frac{A_1}{B}\,\mathbb{E}_{(0)}\left[ j_l \right]+\frac{A_2}{\left(DB \right)^2}\left\{D(D-1)\left(\mathbb{E}_{(0)}\left[ j_l\right]\right)^2+D\, \mathbb{E}_{(0)}\left[ j_l^2\right]\right\}+\frac{A_3}{B^2}\,\mathbb{E}_{(0)}\left[ j_l^2 \right]\nonumber\\
&=A_1X+A_2\left[X^2+\frac{X(1-X)}{DB}\right]+A_3\left[X^2+\frac{X(1-X)}{B}\right]\,.
\end{align}
Using this expression, Eq.~\eqref{new_backward_SI} becomes
\begin{align}
    A_1X + A_2X^2 + A_3 \left \langle x_i^2\right \rangle &=  \left(A_1+\frac{A_2}{DB} + \frac{A_3}{B} + C_1\right) X + \left(A_2-\frac{A_2}{DB}+A_3-\frac{A_3}{B} + C_2\right) X^2  + C_3 \left \langle x_i^2\right \rangle \,.
\end{align}
As this identity must hold for any initial composition of the demes, i.e.\ for any $\bm{i}$, we can identify the terms in $X$, $X^2$ and $\left \langle x_i^2\right \rangle$. While this leads to $3$ equations for our $3$ unknowns $A_1$, $A_2$ and $A_3$, they are not linearly independent. An additional constraint on $A_1$, $A_2$ and $A_3$ stems from the case where the population initially only comprises cooperators, i.e.\ $\bm{i}=(B,\dots,B)\equiv\bm{B}$. In that case, we have $p_{\bm{B}} = 1 + w v_{\bm{B}} = 1$, which gives:
\begin{equation}
v_{\bm{B}} = A_1+A_2+A_3 =0\,.\;
\end{equation}
Solving the resulting set of linear equations yields
\begin{align}
    &A_1 = DB C_1 + (D-1)C_3\,, \\
    &A_2 = -DB\left(C_1+\frac{C_3}{B}\right) \,, \\
    &A_3 = C_3 \,. 
\end{align}

Using Eq.~\eqref{fixation_probability_expansion_SI}, Eq.~\eqref{form} and the above expressions of $A_1$, $A_2$ and $A_3$ provides the explicit expression of the fixation probability:
\begin{align}
    p_{\bm{i}}  &=  X + w \left[(DBC_1+DC_3 )X(1-X) + C_3 \left(\left \langle x_i^2\right \rangle - X\right)\right]  + \mathcal{O}\left(w^2\right) ,
\end{align}
Finally, using Eq.~\eqref{effective_fitness_SI} and Eq.~\eqref{Cdecompo}, and recalling that $\Delta X = wC + \mathcal{O}\left(w^2\right)$, we express the result as a function of the effective fitness parameter $s$ defined in Eq.~\eqref{effective_fitness_SI}, and of the variation $\Delta X$ of overall cooperator fraction during the growth phase:
\begin{align}
    p_{\bm{i}} &=  X + wC + wsDBX(1-X)  + \mathcal{O}\left(w^2\right) \nonumber\\
    &=  X + \Delta X + wsDBX(1-X)  + \mathcal{O}\left(w^2\right) \; . \label{final_formula}
\end{align}
This result shows that the fixation probability of cooperation depends both on the dynamics of the first growth phase, via $\Delta X$, given by Eq.~\eqref{Delta_X_Final_SI}, and on the following cycles of dilution-growth-merging, via the effective fitness parameter $s$. 

\paragraph{Note: fixation of defectors.} The fixation probability $p_{\bm{i}}^{(W)}$ of defectors, starting from $\bm{i} =(i_1,\dots,i_D)$ defectors in each deme, can be obtained by noticing that $p_{\bm{i}}^{(W)}= 1-p_{\bm{i}}$. Denoting by $Y=1-X$ the initial fraction of defectors in the population and by $y_i=1-x_i$ the initial fraction of defectors in each deme, we have $\langle y_i(1-y_i)\rangle=\langle x_i(1-x_i)\rangle$, $\mathrm{Var}(y_i)=\mathrm{Var}(x_i)$ and $Y(1-Y)=X(1-X)$. This yields:
\begin{equation}
    p_{\bm{i}}^{(W)} = Y + \Delta Y -wsDBY(1-Y) + \mathcal{O}\left(w^2\right) \, ,
\end{equation}
with $\Delta Y = - \Delta X$, where $\Delta X$ is given by Eq.~\eqref{Delta_X_Final_SI}. This formula has the same form as Eq.~\eqref{final_formula} for cooperators, but the effective fitness coefficient is $-s$ instead of $s$. 

\subsubsection{Condition on the benefit-to-cost ratio for cooperator fixation to be favored} \label{1.4.1}

The fixation of cooperators is more likely than that of neutral mutants when $p_{\bm{i}} > X$, where $p_{\bm{i}}$ is given by Eq.~\eqref{final_formula} and $X$ is the initial overall fraction of cooperators in the population.  The threshold value of the benefit-to-cost ratio $b/c$ beyond which cooperation is favored is thus obtained by solving $p_{\bm{i}} = X$, which yields:
\small
\begin{equation}
   \frac{b}{c} = \frac{\left[\left \langle x_i(1-x_i)\right \rangle + D(B-1)X(1-X)\right]\left[t-\log\left(\frac{B(e^{t}-1)+K}{K}\right)\right] +  \left[\mathrm{Var}(x_i)+ (D-1) X(1-X)\right]\frac{t}{1+\frac{B}{K-B}e^t}}{ \left[\left \langle x_i(1-x_i)\right \rangle + D(B-1) X(1-X)\right] \left[t-\log\left(\frac{B(K-1)e^{t}+B-K}{K(B-1)}\right)\right] +  \left[\mathrm{Var}(x_i)+(D-1) X(1-X)\right] \frac{t}{1+\frac{B}{K-B}e^t}}\,. \label{Condition_Finale} 
\end{equation}
\normalsize
Eq.~\eqref{Condition_Finale} is used to show the benefit-to cost ratio threshold in Fig. \ref{Fixation_Probability_Deterministic} in the main text.

  \subsection{Comparison of different conditions for cooperation to be favored}

So far, we derived three thresholds of the ratio $b/c$ beyond which cooperation is favored:
\begin{itemize}
    \item Eq.~\eqref{Condition_Growth} determines a condition for cooperation to be advantaged during a specific growth phase of the structured population. It depends on the composition of each deme at the beginning of this growth phase.
    \item Eq.~\eqref{Condition_Cycle} determines a condition for cooperation to be advantaged on average during a dilution-growth-merging cycle. It is obtained by averaging over realizations of the dilution step, and thus over the initial composition of each deme for the growth phase.
    \item Eq.~\eqref{Condition_Finale} determines a condition for cooperator fixation to be advantaged, i.e.\ for cooperators to have a higher fixation probability than neutral mutants. This threshold depends on the initial composition of each deme at the beginning of the whole process, which represents the initialization of the system (how many mutants are initially present in each deme, before a first growth phase and then many cycles of dilution-growth-merging are performed).
\end{itemize}

We expect these conditions to be connected. Indeed, if cooperation is favored during the first growth phase, and then also on average during dilution-growth-merging cycles, for any initial composition of the demes, then it should be favored overall, including with respect to fixation. 

Consider a single initial mutant placed in one deme. In that case, we have:
\begin{align}
    \langle x_i
     \rangle = X= \dfrac{1}{D} \sum_{i=1}^D x_i = \dfrac{1}{DB} \quad \text{and} \quad \langle x_i^2
     \rangle = \dfrac{1}{D} \sum_{i=1}^D x_i^2=\dfrac{1}{DB^2}\,.
\end{align}
Plugging these expressions into Eq.~\eqref{Condition_Finale} yields the same condition as Eq.~\eqref{Condition_Cycle}. Moreover, plugging these expressions into Eq.~\eqref{Condition_Growth} also yields the exact same condition. Thus, we find that the condition for a single cooperator to have a higher fixation probability than a neutral mutant is the same as for cooperation to be favored on average in one cycle of dilution-growth-merging (for any configuration of the demes at the beginning of the growth phase), and as for cooperation to be favored over the first growth phase. Note that this is a particular initial population composition, which is unfavorable for mutants, as the variance in initial deme composition is small. However, it is of interest for fixation, because mutants usually appear in one single individual.

\section{Additional results} \label{3}

\subsection{Exponential growth} \label{3.1}

\subsubsection{Analysis of the growth phase in the exponential case}

Here, we consider our model in the case where the growth phases are exponential  instead of logistic. The exponential growth results below can be obtained from the logistic ones in the main text by taking the limit $K \to \infty$. In this particular case, the absence of saturation of population growth promotes cooperation more broadly, which is why we discuss it explicitly here. 

Under exponential growth, solving the dynamics inside one deme in the weak selection approximation yields
\begin{align}
    N(\tau) &= 
    B e^\tau \left\{1 + w  \tau \left[(b-c)x_0-1\right]\right\} + \mathcal{O}(w^2) \,, \label{sol_N_exp} \\
    x(\tau) &= x_0-w x_0(1-x_0)\left[\tau c +b\log\left(\frac{B-e^{-\tau}}{B-1}\right)\right]  + \mathcal{O}(w^2) \,. \label{sol_x_exp} 
\end{align}
This matches the logistic result in Eq.~\eqref{approximation} in the limit $K \to \infty$. The excess total fraction of cooperators in the full population $\Delta X(\tau)= X(\tau)- X(0)$ during growth then reads, for $\tau\in[0,t]$, 
\begin{align}
    \Delta X(\tau) &= w\left[\Delta X_a (\tau) - \Delta X_b (\tau)\right] + \mathcal{O}\left(w^2\right) \label{Delta_X_Final_Exponential} ,\;\;\textrm{with}\\
    \Delta X_a (\tau) &= \mathrm{Var}(x_i) \tau(b-c) \; ,\;\;\textrm{and}\\
    \Delta X_b(\tau) &= \left \langle x_i(1-x_i)\right \rangle \left[\tau c+b\log\left(\frac{B-e^{-\tau}}{B-1}\right)\right]. 
\end{align}
This matches the logistic result in Eq.~\eqref{Delta_X_Final_SI} in the limit $K \to \infty$. 

\begin{figure}[H]
    \centering
    \includegraphics[width=0.6\textwidth]{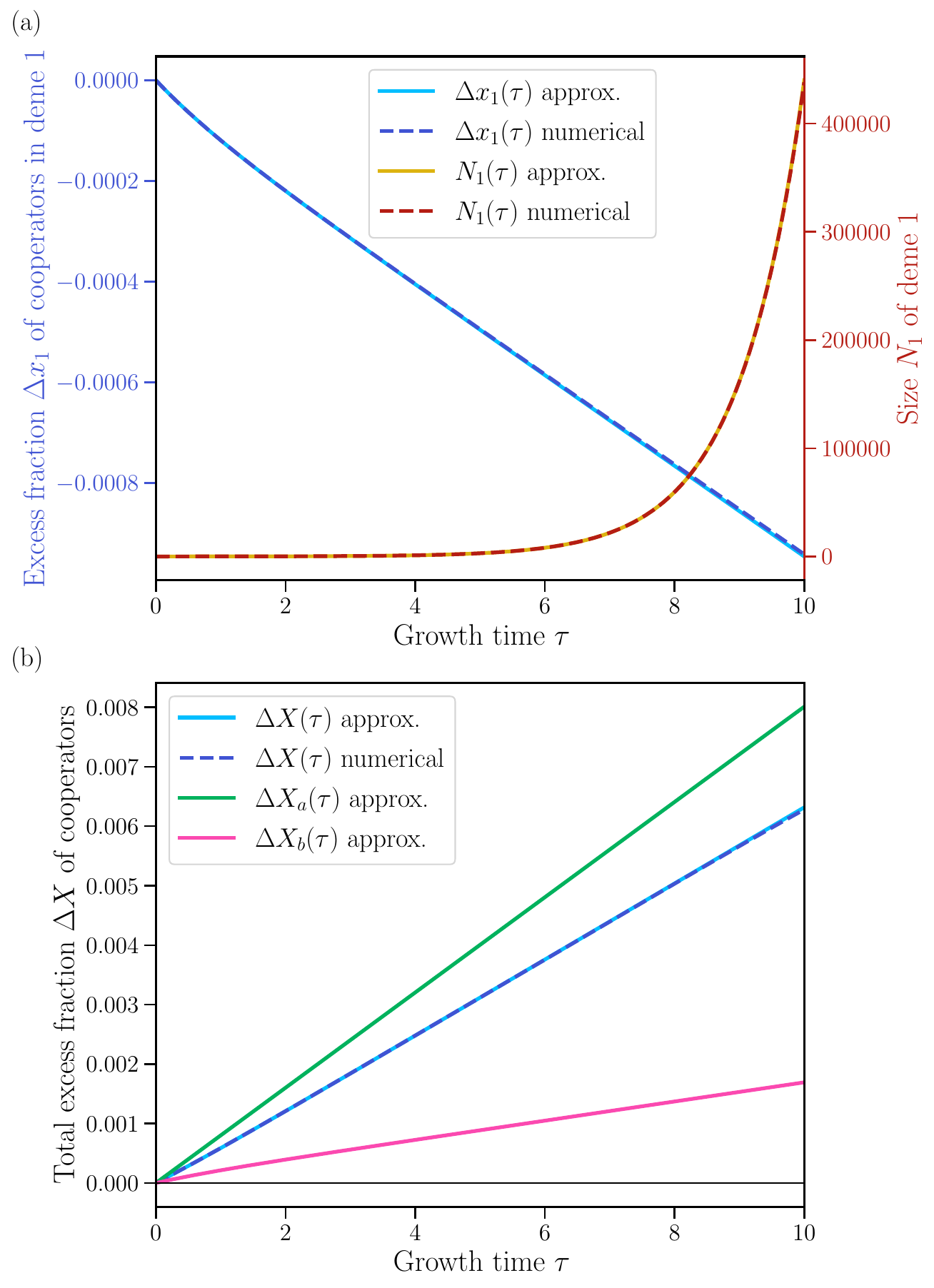}
    \caption{\textbf{Exponential growth phase: evolution of the fraction of cooperators in a deme and in the population.} Same as Fig.~\ref{Growth_Phase} in the main text, but with exponential growth. \textbf{(a)} Excess fraction $\Delta x_1 (\tau)= x_1(\tau) - x_1(0)$ of cooperators in one deme, namely deme $1$, and size $N_1(\tau)$ of this deme, versus growth time $\tau$. We show the numerical solution of Eq.~\eqref{equation_diff} for $K=10^8$ and our analytical approximation of Eq.~\eqref{sol_x_exp} and Eq.~\eqref{sol_N_exp}, obtained in the weak selection regime.
    \textbf{(b)} Excess fraction $\Delta X = X(\tau) - X(0)$ of cooperators in the total population versus growth time $\tau$. As in (a), we show the numerical solution and the analytical approximation in the weak selection regime. Here, the latter is given by Eq.~\eqref{Delta_X_Final_Exponential}. In addition, we show the two terms into which we decompose  Eq.~\eqref{Delta_X_Final_Exponential}.
    As in Fig.~\ref{Growth_Phase} in the main text, results are obtained for a population of $D=5$ demes, each with initial size $B=20$, using initial compositions $x_1(\tau=0)=1/10$, $x_2(0)=1/4$, $x_3(0)=1/2$, $x_4(0)=3/4$ and $x_5(0)=9/10$, fitness intensity $w=10^{-3}$, benefit $b=10$ and cost $c=1$. 
    }
    \label{Growth_Phase_Exponential}
\end{figure}

Fig. \ref{Growth_Phase_Exponential} shows the time evolution of cooperator fraction in a deme and in the whole population, as well as the time evolution of deme size, during an exponential growth phase. As in Fig.~\ref{Growth_Phase} in the main text, which shows this in the case of logistic growth, we observe a very good agreement between numerical results and our analytical approximation, which is here given by Eq.~\eqref{Delta_X_Final_Exponential}. Comparing Fig. \ref{Growth_Phase_Exponential} with Fig.~\ref{Growth_Phase} in the main text reveals a key difference between exponential and logistic growth. With exponential growth, when cooperation is promoted during growth at the scale of the full population, the effect is not transient, but increases over time. This is because, in the absence of saturation of deme size, the difference between the sizes of demes with a high and a smaller fraction of cooperators keeps increasing during growth.

\subsubsection{Condition on the benefit-to-cost ratio for cooperation to be favored during growth} \label{3.1.2}

As with logistic growth (see Section~\ref{1.2.3}), the threshold value of the benefit-to-cost ratio beyond which cooperation is favored during a growth phase is obtained by solving $\Delta X (t)= 0$. It yields:
\begin{equation}
    \frac{b}{c} =\frac{ \left[\langle x_i(1-x_i) \rangle  + \mathrm{Var}(x_i)\right] t }{\mathrm{Var}(x_i) \, t- \left \langle x_i(1-x_i)\right \rangle \log\left(\frac{B-e^{-t}}{B-1}\right) }\,. \label{Condition_Growth_Exponential} 
\end{equation}
In particular, in the limit $t\to\infty$, Eq.~\eqref{Condition_Growth_Exponential} reduces to
\begin{equation}
    \frac{b}{c} =1+\frac{\langle x_i(1-x_i) \rangle}{\mathrm{Var}(x_i)} = \frac{\langle x_i \rangle - \langle x_i \rangle ^2}{\mathrm{Var}(x_i)}\,. 
    \label{Condition_Growth_Exponential_Large_Time} 
\end{equation}

\subsubsection{Condition on the benefit-to-cost ratio for cooperation to be favored on average over a cycle of dilution-growth-merging}

As with logistic growth (see Section~\ref{1.3.1}), the threshold value of the benefit-to-cost ratio beyond which cooperation is favored on average over a cycle of dilution-growth-merging is obtained by solving $ \mathbb{E}\left[\Delta X^\prime\right] = \mathbb{E}\left[\Delta X(t)\right] >0$. With exponential growth, we have $ \mathbb{E}\left[\Delta X^\prime\right] = w s X_n^\prime\left(1-X_n^\prime\right) + \mathcal{O}\left(w^2\right)$ with
\begin{equation}
    s  = \frac{D-1}{DB}t(b-c)  -\frac{B-1}{B}\left[tc+b\log\left(\frac{B-e^{-t}}{B-1}\right)\right]\,. \label{effective_fitness_SI_exp} 
\end{equation}
This leads to the threshold value
\begin{equation}
    \frac{b}{c} = \frac{(DB-1)t}{(D-1) t-(B-1)D\log\left(\frac{B-e^{-t}}{B-1}\right)  }\,. \label{Condition_Cycle_exp} 
\end{equation}
In particular, in the limit $t\to\infty$, Eq.~\eqref{Condition_Cycle_exp} reduces to
\begin{equation}
    \frac{b}{c} = \frac{DB-1}{D-1}\,. \label{Condition_Cycle_exp_Large_Time} 
\end{equation}

\subsubsection{Condition on the benefit-to-cost ratio for cooperator fixation to be favored}

As with logistic growth (see Section~\ref{1.4.1}), the threshold value of the benefit-to-cost ratio $b/c$ for the fixation of a single cooperator to be more favored than that of a neutral mutant is obtained by solving $p_{\bm{i}} = X$. With exponential growth, this yields:
\begin{equation}
   \frac{b}{c} = \frac{\left[\mathrm{Var}(x_i) + \left \langle x_i(1-x_i)\right \rangle + (DB-1)X(1-X)\right]\,t}{ \left[\mathrm{Var}(x_i)+(D-1) X(1-X)\right] t-\left[\left \langle x_i(1-x_i)\right \rangle + D(B-1) X(1-X)\right] \log\left(\frac{B-e^{-t}}{B-1}\right) }\,. \label{Condition_Finale_Exponential} 
\end{equation}
In particular, in the limit $t\to\infty$, Eq.~\eqref{Condition_Finale_Exponential} reduces to:
\begin{equation}
   \frac{b}{c} =  \frac{\mathrm{Var}(x_i)+\left \langle x_i(1-x_i)\right \rangle + (DB-1)X(1-X)}{\mathrm{Var}(x_i)+(D-1) X(1-X)} 
   \,. \label{Condition_Finale_Exponential_Large_Time} 
\end{equation}

\subsection{Analysis of some extreme parameter regimes} \label{extrreg}

We showed in Section~\ref{efffit} that $\mathbb{E}\left[\Delta X^\prime\right] = w s X_n^\prime\left(1-X_n^\prime\right) + \mathcal{O}\left(w^2\right)$, where the effective fitness parameter $s$ is given by Eq.~\eqref{effective_fitness_SI}. 
This equation was derived for $w\ll 1$, with $b>c$, $B<K$. So far, we considered by default that all parameters, including $B,\, K,\, t$, were of order unity (in $w$). 

In this Section, we analyze the impact of different limits of the parameters $B,\, K$ and $t$ on the effective fitness $s$. This allows to understand whether cooperation can be favored in these extreme regimes. Indeed, we showed above that cooperation is favored on average over one cycle of dilution-growth-merging if and
only if $s>0$.

\subsubsection{Large growth time $t$: saturating demes} \label{larget}
Let us assume that the growth time $t$ is large, specifically with the scaling $t=\log(\alpha/w)$, where $\alpha>0$ is a constant of order unity. Then, $e^t$ scales as $1/w$. 
Here, we assume that $K$ is of order unity (in terms of $w$).
Eq.~\eqref{effective_fitness_SI} becomes:
\begin{align}
    s&=\frac{D-1}{DB}\frac{(b-c)\log(\alpha/w)}{1+\frac{B}{K-B}\frac{\alpha}{w}}  \nonumber\\&-\frac{B-1}{B}\left[(c-b)\log\left(\frac{\alpha}{w}\right)+b\log\left(\frac{B(K-1)\frac{\alpha}{w}+B-K}{K(B-1)}\right)-c \log\left(\frac{B(\frac{\alpha}{w}-1)+K}{K}\right)\right].
\end{align}
With $w \ll 1$, the first term reads:
\begin{align}
    \dfrac{D-1}{DB}\frac{(b-c)\log(\alpha/w)}{1+\frac{B}{K-B}\frac{\alpha}{w}} =\dfrac{D-1}{DB}\dfrac{K-B}{B}(b-c)\frac{w}{\alpha}\log\left(\frac{\alpha}{w}\right)\left[1 +\mathcal{O}(w)\right]=\mathcal{O}(w)\,.
\end{align}
Within the second term, we have:
\begin{align}
    \log\left(\dfrac{\alpha}{w}\right)&(c-b)+b\log\left(\frac{B(K-1)\frac{\alpha}{w}+B-K}{K(B-1)}\right)-c \log\left(\frac{B(\frac{\alpha}{w}-1)+K}{K}\right)\\
    &= c \log \left( \dfrac{K}{B } \right) +b \log \left(\dfrac{B(K-1)}{K(B-1)}\right) + \mathcal{O}(w).
\end{align}
We note that
\begin{equation}
    c \log \left( \dfrac{K}{B } \right) +b \log \left(\dfrac{B(K-1)}{K(B-1)}\right) = c \log \left(\dfrac{K-1}{B-1}\right) + (b-c) \log\left(\dfrac{B}{K} \dfrac{K-1}{B-1}\right)>0\,.
\end{equation}
Indeed, since $B<K$, we have $B(K-1)>K(B-1)$. Since in addition $b>c$, the expression above is positive.
Therefore, we obtain
\begin{equation}
    s=-\dfrac{B-1}{B} \left[ c \log \left( \dfrac{K}{B} \right) +b \log \left(\dfrac{B(K-1)}{K(B-1)}\right)\right] +\mathcal{O}(w)<0 \,.
\end{equation}

Thus, in the regime of large growth time $t$, with $t$ scaling as $\log(1/w)$, cooperation cannot be advantaged in the population. Indeed, qualitatively, when $t$ is large enough, all demes reach saturation during logistic growth. In our model, they all have the same size $K$  (which is of order unity in terms of $w$) at the end of growth. Hence, demes with larger cooperator fractions lose their growth advantage and cooperation is not advantaged. 

Besides, our result is consistent with Figure~\ref{Growth_Phase}(b), where the excess fraction of cooperators $\Delta X$ becomes negative when the growth time $\tau$ reaches $\tau_0 \simeq 5.5$, while $\log(1/w) \simeq 7$.

\subsubsection{Large carrying capacity $K$: exponential growth} \label{largeK}
In the limit $K\rightarrow\infty$, we recover the case of exponential growth that was covered in Section~\ref{3.1}. In particular, Eq.~\eqref{effective_fitness_SI} then reduces to Eq.~\eqref{effective_fitness_SI_exp}.

\subsubsection{Large bottleneck size $B$} \label{largeB}
In the limit $B\rightarrow \infty $, we also have $K \rightarrow \infty$, as $K>B$. Thus, we can start from Eq.~\eqref{effective_fitness_SI_exp}. When $B\rightarrow \infty $, it becomes to leading order
\begin{equation}
    s = -tc \,.
\end{equation}

This effective fitness is always negative, and cooperation cannot be advantageous. Qualitatively, small sizes are necessary to maintain contrast across demes after the dilution step, which is a key ingredient favoring cooperation. 

Note that the expression of $s$ obtained in this limit is reminiscent of the effective fitness obtained in~\cite{Moawad_2023}. In that work, demes have large size, and cooperation is never favored. This is because demes are large, but also because selection is soft (see main text). 

\section{Simulation method for stochastic growth} \label{2}

To incorporate stochastic growth, we change our model by assuming that individuals can divide with birth rates given by $k_{M,i}^{+}=f_M\left[1-\left(N_{W, i}+N_{M, i}\right) / K\right]$ for mutants and $k_{W,i}^{+}=f_W\left[1-\left(N_{W, i}+N_{M, i}\right) / K\right]$ for wild-type individuals. Here, $N_{M, i}$ and $N_{W, i}$ denote the numbers of mutant and wild-type individuals in deme $i$. This means that $N_{M, i}$ and $N_{W, i}$ now follow a master equation, in contrast to the deterministic case that follows ordinary differential equations (see Eq.~\eqref{approximation}).

To simulate stochastic growth in a deme, we use the Gillespie algorithm \cite{gillespie1976general,gillespie1977exact} which does not require any time discretization. In a metapopulation of $D$ demes, the possible `reactions' for the two types $M$ and $W$ in each deme $i$ are:
\begin{itemize}
    \item $M_i \xrightarrow{k_{M,i}^{+}} M_i+M_i$ : birth of one mutant in deme $i$ with rate $k_{M,i}^{+}=f_M\left[1-\left(N_{W, i}+N_{M, i}\right) / K\right]$,
    \item $W_i \xrightarrow{k_{W,i}^{+}} W_i+W_i$ : birth of one wild-type in deme $i$ with rate $k_{W,i}^{+}=f_W\left[1-\left(N_{W, i}+N_{M, i}\right) / K\right] $.
\end{itemize}
Note that we do not consider any death event in our model. Then, we define the total rate of possible events as
\begin{equation}
k_{\text{tot}}=\sum_{i=1}^D\left(k_{M, i}^{+} N_{M, i} + k_{W, i}^{+} N_{W, i}\right)\,.
\end{equation}
During growth, i.e.\ for $\tau \in[0, t]$, birth events occur at time intervals drawn in an exponential distribution with mean $1 / k_{\text{tot}}$, and the specific reaction that occurs is selected randomly proportionally to the ratio of its rate and $k_{\text{tot}}$. The merging and dilution step are implemented exactly as with deterministic growth. 

\end{document}